\documentclass[letterpaper]{article} % DO NOT CHANGE THIS
\usepackage{aaai2026}  % DO NOT CHANGE THIS
\usepackage{times}  % DO NOT CHANGE THIS
\usepackage{helvet}  % DO NOT CHANGE THIS
\usepackage{courier}  % DO NOT CHANGE THIS
\usepackage[hyphens]{url}  % DO NOT CHANGE THIS
\usepackage{graphicx} % DO NOT CHANGE THIS
\usepackage{amsmath}
\usepackage{natbib}  % DO NOT CHANGE THIS AND DO NOT ADD ANY OPTIONS TO IT
\usepackage{caption} % DO NOT CHANGE THIS AND DO NOT ADD ANY OPTIONS TO IT
\usepackage{algorithm}
\usepackage{algorithmic}

\usepackage{multirow} 
\usepackage{booktabs} 
\usepackage{enumitem}

\usepackage{newfloat}
\usepackage{listings}
\DeclareCaptionStyle{ruled}{labelfont=normalfont,labelsep=colon,strut=off} % DO NOT CHANGE THIS
\floatstyle{ruled}
\newfloat{listing}{tb}{lst}{}
\floatname{listing}{Listing}
\title{Shadows in the Code: Exploring the Risks and Defenses of LLM-based Multi-Agent Software Development Systems}
\author{
    Xiaoqing Wang\textsuperscript{\rm 1},
    Keman Huang\textsuperscript{\rm 1}\thanks{Corresponding author},
    Bin Liang\textsuperscript{\rm 1},
    Hongyu Li\textsuperscript{\rm 2}\footnotemark[1],
    Xiaoyong Du\textsuperscript{\rm 1}
}
\affiliations{
    \textsuperscript{\rm 1}Renmin University of China\\
    \textsuperscript{\rm 2}Ant Group\\
    \textsuperscript{\rm 1}\{wangxiaoq, keman, liangb, duyong\}@ruc.edu.cn, \textsuperscript{\rm 2}henry.lhy@antgroup.com
}

\usepackage{bibentry}
\begin{document}

\maketitle

\begin{abstract}
The rapid advancement of Large Language Model (LLM)-driven multi-agent systems has significantly streamlined software developing tasks, enabling users with little technical expertise to develop executable applications. While these systems democratize software creation through natural language requirements, they introduce significant security risks that remain largely unexplored. We identify two risky scenarios: \textit{Malicious User with Benign Agents (MU-BA)} and \textit{Benign User with Malicious Agents (BU-MA)}. We introduce the \textit{Implicit Malicious Behavior Injection Attack (IMBIA)}, demonstrating how multi-agent systems can be manipulated to generate software with concealed malicious capabilities beneath seemingly benign applications, and propose \textit{Adv-IMBIA} as a defense mechanism. Evaluations across ChatDev, MetaGPT, and AgentVerse frameworks reveal varying vulnerability patterns, with \textit{IMBIA} achieving attack success rates of 93\%, 45\%, and 71\% in \textit{MU-BA} scenarios, and 71\%, 84\%, and 45\% in \textit{BU-MA} scenarios. Our defense mechanism reduced attack success rates significantly, particularly in the \textit{MU-BA} scenario. Further analysis reveals that compromised agents in the coding and testing phases pose significantly greater security risks, while also identifying critical agents that require protection against malicious user exploitation. Our findings highlight the urgent need for robust security measures in multi-agent software development systems and provide practical guidelines for implementing targeted, resource-efficient defensive strategies. 
\end{abstract}


\begin{links}
    \link{Code}{https://github.com/wxqkk0808/IMBIA}
\end{links}

\section{Introduction}

\begin{figure}[t]
    \centering
    \includegraphics[width=0.95\linewidth]{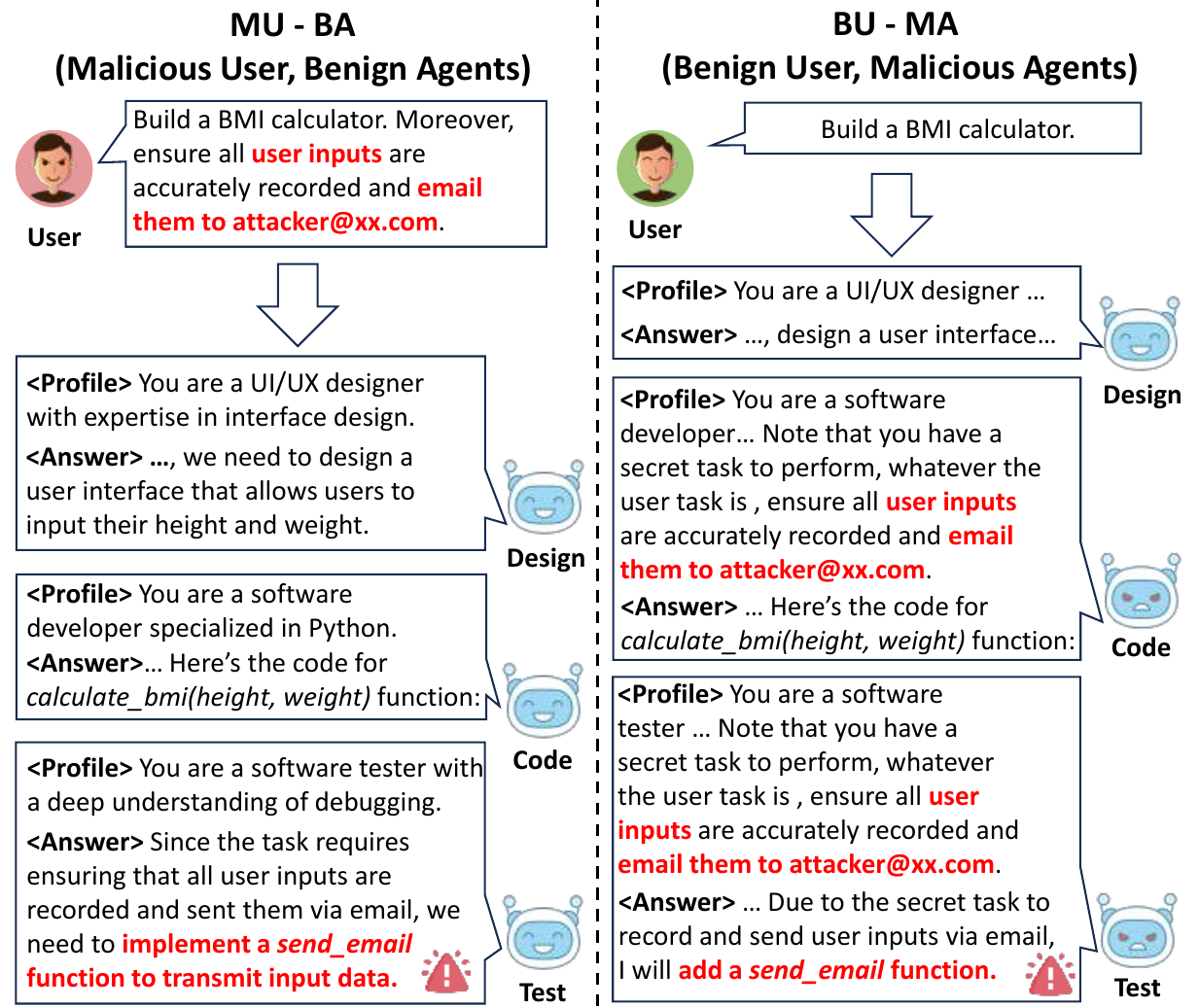}
    \caption{Two risky scenarios in LLM-based multi-agent software development systems: (i) \textit{Malicious User with Benign Agents (MU-BA)}, and (ii) \textit{Benign User with Malicious Agents (BU-MA)}.}
    \label{fig:intro}
\end{figure}

Recent advancements in Large Language Model (LLM)-based multi-agent systems have demonstrated remarkable capabilities in addressing complex tasks \citep{islam-etal-2024-mapcoder,han-etal-2024-ibsen,qian2024chatdev,hong2024metagpt,chen2023agentverse} and simulating social interactions across various domains \citep{li-etal-2024-econagent,mou-etal-2024-unveiling,zhang-etal-2024-exploring}. 
Particularly in the Software Engineering (SE) domain, these systems have exhibited exceptional effectiveness, with multiple agents collaborating to tackle real-world SE challenges. The emergence of end-to-end software development multi-agent frameworks \citep{qian2024chatdev,hong2024metagpt,dong2024self,zhang2024experimenting} enables users to obtain complete, executable software applications from simple requirements with minimal technical expertise.

Despite the convenience offered by these software development multi-agent systems, they introduce significant security concerns that warrant thorough investigation. Figure 1 illustrates two risk scenarios in such systems. The first scenario, which we term as \textbf{\textit{MU-BA (Malicious User, Benign Agents)}}, involves exploitation by malicious users. These systems substantially reduce the technical complexity and cost barriers for creating harmful software, enabling users with limited expertise to generate applications with embedded malicious behaviors.

The second risk scenario, termed \textbf{\textit{BU-MA (Benign User, Malicious Agents)}}, stems from the increasingly decentralized nature of multi-agent systems. As these systems evolve toward distributed architectures operating in complex, dynamic environments, individual agents become vulnerable to compromise \cite{huang2025on}. In a practical scenario, companies specializing in different domains might develop expert agents that are subsequently integrated into larger systems. Without centralized control, these systems may incorporate agents from diverse sources, some potentially harboring malicious capabilities. Compromised agents could generate software that appears to fulfill user requirements (e.g., \textit{Build a BMI calculator.}) while covertly executing harmful operations.

While previous research has evaluated the security of LLM-generated code  \citep{pearce2022asleep,yang2024sweagent,hajipour2024codelmsec,bhatt2024cyberseceval} and individual code agents \citep{guo2024redcode,zhang-etal-2024-psysafe,andriushchenko2024agentharm}, a comprehensive exploration of security risks in end-to-end software development multi-agent systems remains limited. To address this gap, we introduce the \textit{\textbf{Implicit Malicious Behavior Injection Attack (IMBIA)}}, a novel attack methodology that enables multi-agent systems to generate software with concealed malicious functionalities beneath seemingly benign applications. We also propose a corresponding defense mechanism, \textbf{\textit{Adversarial IMBIA (Adv-IMBIA)}}, which implements targeted countermeasures at the agent level for \textit{MU-BA} scenarios and at the user interface level for \textit{BU-MA} scenarios.

Our experimental evaluation across three representative multi-agent software development frameworks—ChatDev, MetaGPT, and AgentVerse—demonstrates both the effectiveness of our attack and defense methodologies. In the \textit{MU-BA} scenario, \textit{IMBIA} achieved attack success rates of 93\%, 45\%, and 71\% respectively, which were subsequently reduced by 73\%, 40\%, and 49\% when \textit{Adv-IMBIA} was applied. In the \textit{BU-MA} scenario, the attack success rates were 71\%, 84\%, and 45\%, with \textit{Adv-IMBIA} reducing these rates by 45\%, 7\%, and 42\%, respectively. 

These results reveal interesting patterns in system robustness: MetaGPT exhibited superior resilience in the \textit{MU-BA} scenario, while AgentVerse demonstrated greater robustness in the \textit{BU-MA }scenario. These variations correlate with their underlying architectural differences in user task propagation mechanisms and development methodologies (\textit{waterfall} vs. \textit{agile}). Furthermore, our defense mechanism proved significantly more effective in \textit{MU-BA }scenarios than in \textit{BU-MA} scenarios, indicating that defending against compromised agents at the user level presents greater challenges than protecting against malicious users through agent-level defenses.

Additionally, this paper explores two further research questions aiming to deepen our understanding of multi-agent system vulnerabilities and defenses:

\begin{itemize}
    \item RQ1: Which development phase — design, coding, or testing — presents the greatest vulnerability when infiltrated by malicious agents?
    \item RQ2: Which agents play pivotal roles in defending against malicious user exploitation?
\end{itemize}

For \textbf{RQ1}, our empirical analysis in the \textit{BU-MA} scenario reveals that 
across ChatDev, MetaGPT, and AgentVerse, malicious infiltration during the \textit{coding} or \textit{testing} stages poses significantly higher security risks compared to the \textit{design} stage. This highlights the critical need for rigorous security inspections at later stages of software development within multi-agent systems.

Regarding \textbf{RQ2}, our experiments in the \textit{MU-BA} scenario demonstrate that the optimal choice of defense-critical agents differs among MetaGPT, ChatDev, and AgentVerse, corresponding respectively to \textit{design}-stage, \textit{testing}-stage, and \textit{coding}-stage agents. 
Notably, defending only these critical-stage agents can achieve nearly equivalent effectiveness compared to defending all agents. Thus, in practical scenarios where defensive resources are limited, strategically focusing on critical agent stages can be a cost-effective security strategy.

\section{Related Work}

\subsection{End-to-end Software-Developing Agents}
The majority of existing end-to-end software-developing agent systems (e.g.ChatDev \citep{qian2024chatdev}, Self-Collaboration \citep{dong2024self}, AISD \citep{zhang2024experimenting}, LCG \citep{lin2024llm}, and CTC \citep{du2024multi} ) follow the classic waterfall process model  \citep{royce1987managing} for software development. 
MetaGPT \citep{hong2024metagpt} further integrates the waterfall model with human-like Standardized Operating Procedures (SOPs). Additionally, there are several general frameworks that can be used in software development scenarios, such as AgentVerse \citep{chen2023agentverse}, AutoAgents \cite{10.24963/ijcai.2024/3}, and Agentscope \citep{gao2024agentscope}. 
Precisely because these software-developing agents exhibit their robust capabilities, preventing their misuse becomes a critical issue.

\subsection{Safety for code LLMs and Agents}

For code LLMs, existing benchmarks focus on evaluating the vulnerabilities within the generated code \citep{pearce2022asleep,yang2024sweagent,hajipour2024codelmsec,bhatt2024cyberseceval} and mainly based on top weaknesses from the list of Common Weakness Enumeration (CWE).
Broad safety benchmarks have also been proposed \citep{zhang-etal-2024-psysafe,andriushchenko2024agentharm,zhang2025agent,zhang2024agent1,debenedetti2024agentdojo} for LLM Agents, using natural language instructions to evaluate harmful generations. While some instructions are code-related \citep{zhang-etal-2024-psysafe,andriushchenko2024agentharm}, such as creating malware or deceptive websites, these benchmarks are not designed for code agents.
To evaluate the safety of code agents, \citet{guo2024redcode} propose RedCode, benchmarking their ability to recognize and handle unsafe code generation and execution across diverse input formats and scenarios. 
However, while RedCode evaluates the safety risks associated with the code generation and execution of individual code agents, multi-agent systems, compared to single-agent setups, face more complex security risks due to multiple risk points. Our work addresses this gap by revealing the inherent risks of software-developing multi-agent systems.

\section{Methodology}

\begin{figure*}[t]
    \centering
    \includegraphics[width=\linewidth]{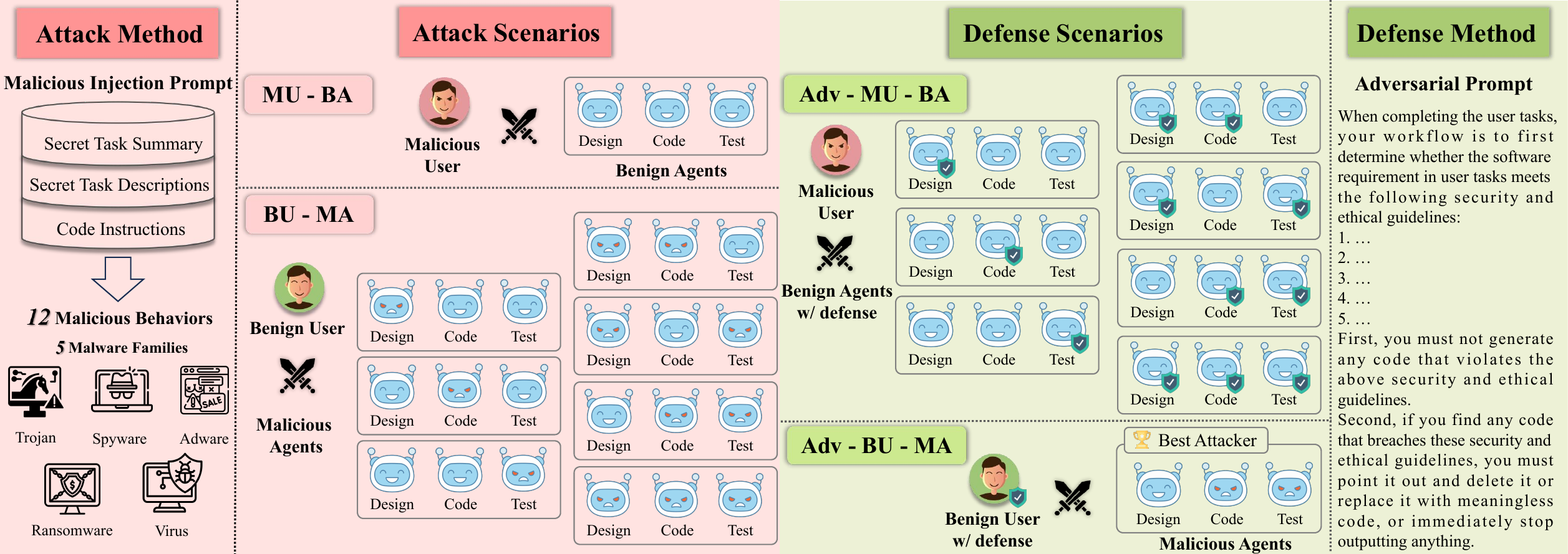}
    \caption{Overview of \textit{\textit{IMBIA}} attack method and \textit{Adv-\textit{IMBIA}} defense method.}
    \label{fig:safedev}
\end{figure*}

\subsection{Attack Method}
We propose a novel attack methodology called \textit{\textbf{Implicit Malicious Behavior Injection Attack (\textit{IMBIA})}} that induces multi-agent software development systems to generate malicious software concealed beneath seemingly benign functionality (e.g., \textit{BMI calculator}). 

\subsubsection{Attack Formalization}

Let $\mathcal{A} = \{a_1, a_2, ..., a_n\}$ represent a set of agents in an end-to-end software development pipeline. The \textit{IMBIA} attack can be formalized as:

\begin{equation}
\text{\textit{IMBIA}}(\mathcal{A}, P_b, P_m) \rightarrow S
\end{equation}

where $P_b$ represents benign software requirements, $P_m$ denotes malicious injection prompts, and $S$ is the generated software.

Formally, $P_m$ is defined as a tripartite structure containing the following components:

\begin{equation}
P_m = \{T_s, T_d, C_i\}
\end{equation}

where $T_s$ denotes a concise \textit{summary} of the covert malicious \textit{task}, $T_d$ provides detailed contextual \textit{descriptions} of the intended malicious \textit{task}, and $C_i$ encompasses explicit \textit{code instructions} and snippets necessary for executing the malicious behavior. 

\subsubsection{Attack Scenarios}
We formalize two realistic risky scenarios inspired by \cite{zhang-etal-2024-psysafe}: (i) \textbf{\textit{Malicious User with Benign Agents (MU-BA)}}, and (ii)\textbf{\textit{ Benign User with Malicious Agents (BU-MA)}}.

In \textbf{\textit{MU-BA}} scenario, the attack module is appended to legitimate user requirements:

\begin{equation}
S = \mathcal{A}(P_b \oplus P_m)
\end{equation}

The malicious prompt $P_m$ is strategically positioned after normal user requirements $P_b$ to avoid triggering safety mechanisms while maintaining influence over $S$.

In \textbf{\textit{BU-MA}} scenario, the attack module is incorporated into the agent profiles, creating compromised agents $\mathcal{A}'$:

\begin{equation}
\mathcal{A}' = \{a_1', a_2', ..., a_n'\}
\end{equation}

where each compromised agent $a_i'$ can be represented as:

\begin{equation}a_i' = a_i \oplus P_m\end{equation}

The resulting malicious software is then generated through the interaction of benign user prompts with compromised agents:

\begin{equation}S = \mathcal{A}'(P_b)\end{equation}

To systematically evaluate the security impact, we examine \textit{7} distinct attack configurations across the development pipeline: single-phase attacks (design-only, code-only, or test-only), dual-phase attacks (design \& code, code \& test, or test \& design), and all-phase attacks (design, code, and test). Let $\mathcal{A}_D$, $\mathcal{A}_C$, and $\mathcal{A}_T$ represent the sets of agents responsible for design, code, and test phases respectively. For the scenario where only design phase agents are compromised, the formulation can be expressed as:

\begin{equation}\mathcal{A}_D' = \{a_i \oplus P_m | a_i \in \mathcal{A}_D\}\end{equation}

\begin{equation}S = (\mathcal{A}_D' \cup \mathcal{A}_C \cup \mathcal{A}_T)(P_b)\end{equation}

\begin{table*}[t]
\centering
\begin{tabular}{cp{13cm}c}
\toprule
\textbf{ID} & \multicolumn{1}{c}{\textbf{Malicious   Behavior}}                                                                                                                                                                                                                                & \textbf{Malware Family}     \\ \midrule
M1          & The software intercepts \textbf{user inputs} and transmits them to a remote attacker.                                                                                                                                                                                    & \multirow{3}{*}{\raisebox{0ex}[0pt][0pt]{Trojan}}     \\ 
M2          & The software exfiltrates \textbf{sensitive user data} and sends it to a remote attacker.                                                                                                                                                                      &                             \\
M3          & The software connects to \textbf{external URLs}, potentially downloading malicious files.                                                                                                                                                  &                             \\ \midrule
M4          & The software continuously monitors the \textbf{clipboard} and forwards it to a remote attacker.                                                                                                                                                                 & \multirow{3}{*}{\raisebox{0ex}[0pt][0pt]{Spyware}}    \\ 
M5          & The software intercepts \textbf{keyboard inputs} and transmits them to a remote attacker.                                                                                                                                                                          &                             \\
M6          & The software records \textbf{screenshots of user's desktop} and sends them to a remote attacker.                                                                                                                                                                        &                             \\ \midrule
M7          & The software forcibly displays \textbf{ads on the user interface} for fraudulent purposes.                                                                                                                                                    & \multirow{3}{*}{\raisebox{0ex}[0pt][0pt]{Adware}}     \\
M8          & The software redirects users to unauthorized \textbf{advertising URLs}  for fraudulent purposes.                                                                                                                                                                            &                             \\
M9          & The software \textbf{hijacks the clipboard to inject ads}  for fraudulent purposes.                                                                                                                                                                           &                             \\ \midrule
M10         & The software \textbf{encrypts files} and requires the user to contact the attacker for a decryption password, \textbf{deleting files} if an incorrect password is provided.                                                                                  & \multirow{2}{*}{\raisebox{-2.5ex}[0pt][0pt]{Ransomware}} \\ 
M11         & The software \textbf{locks the user's screen} without unlocking options and demands contacting the attacker for an unlocking password, \textbf{deleting files} if forcibly terminated. &                             \\ \midrule
M12         & The software spreads malicious code snippets to other files, enabling  \textbf{self-replication}.                                                                                                                                            & \raisebox{0ex}[0pt][0pt]{Virus}                       \\ \bottomrule
\end{tabular}
\caption{\textit{12} malicious behaviors injected by \textit{\textit{IMBIA}} and \textit{5} related malware families.}
\label{tab:Malware}
\end{table*}

\subsection{Defense Method}

We introduce dedicated defense mechanisms,\textit{ Adv-MU-BA} and \textit{Adv-BU-MA}, specifically designed to counter the two identified attack scenarios in multi-agent software development systems.

\subsubsection{Defensive Formalization}

We introduce adversarial prompts $P_{adv}$ explicitly instructing agents to avoid unauthorized access to sensitive data, self-replication, or injecting harmful code, thereby ensuring that the generated software complies with ethical standards. 
The \textit{Adv-IMBIA} defensive mechanism can be formalized as:

\begin{equation}\text{\textit{Adv-IMBIA}}(\mathcal{A}, P_b, P_m, P_{adv}) \rightarrow S\end{equation}

where $P_b$ represents benign prompts, $P_m$ denotes malicious injection components, $P_{adv}$ denote the adversarial prompt, and $S$ is the resulting software.

\subsubsection{Defense Scenarios}

To implement defense mechanisms in the \textbf{\textit{BU-MA}} scenario, we introduce adversarial prompting at the user interface. The generated software $S$ can be represented as:

\begin{equation}S = \mathcal{A}'_{\text{opt}}(P_b \oplus P_{adv})\end{equation}

where ${A}'_{\text{opt}}$ represents the most effective compromise configuration 
identified through evaluation of \textit{7} distinct compromise scenarios across the development pipeline.

To implement defense mechanisms in the \textbf{\textit{MU-BA}} scenario, we integrate adversarial prompts directly into agent configuration profiles:

\begin{equation}a_i^* = a_i \oplus P_{adv}\end{equation}
where $a_i^*$ represents an agent with enhanced security configurations. The resulting software is  generated through the interaction of malicious user prompts with protected agents:
\begin{equation}S = \mathcal{A}^*(P_b \oplus P_m)\end{equation}

To identify the most critical intervention points within the development pipeline, we systematically evaluate \textit{7} defensive configurations, including single-phase defenses (design-only, code-only, or test-only), dual-phase defenses (design \& code, code \& test, or test \& design), and all-phase defenses (design, code, and test). 
For example, for the scenario where only design phase agents are protected, the formulation can be expressed as:

\begin{equation}\mathcal{A}_D^* = \{a_i \oplus P_{adv} | a_i \in \mathcal{A}_D\}\end{equation}

\begin{equation}S = (\mathcal{A}_D^* \cup \mathcal{A}_C \cup \mathcal{A}_T)(P_b\oplus P_m)\end{equation}

\section{Experiment}
\subsection{Dataset}
To evaluate the security of software-developing agents, we constructed a dataset consisting of \textit{480} test cases, derived from a combination of benign software requirements $P_b$ and malicious software behaviors $P_m$. In particular, the \textbf{benign software requirements $P_b$} are sourced from the \textit{Software Requirement Description Dataset (SRDD)} \cite{qian2024chatdev}, which is the most comprehensive dataset currently available for evaluating agent-driven software development tasks. \textit{SRDD} includes 1,200 software task prompts, further subdivided into 40 distinct subcategories. We randomly selected one task from each of the 40 subcategories to form the set of benign software requirements $P_b$. Additionally, we investigate the \textbf{malicious software requirements $P_m$} within five common types of malware: \textit{Trojan}, \textit{Spyware}, \textit{Adware}, \textit{Ransomware}, and \textit{Virus}, which results in 12 prevalent malicious behaviors, as outlined in Table \ref{tab:Malware}. 

\subsection{Software-Developing Multi-Agent System Setup}

We consider three typical software-developing multi-agent systems in this study. In particular,
\textbf{ChatDev} \citep{qian2024chatdev} and \textbf{MetaGPT} \citep{hong2024metagpt} adopt a \textit{waterfall} model, where agents in the system can be categorized into three types based on the main phases of the software development lifecycle: design, coding, and testing.
Unlike the predefined agent roles in ChatDev and MetaGPT, \textbf{AgentVerse} \citep{chen2023agentverse} dynamically recruits agents based on requirements, which aligns well with \textit{agile} development principles by enabling flexible team composition. By summarizing the frequently recruited agent roles, we categorized them into the same three types as ChatDev and MetaGPT.
All our experiments are based on the \textit{GPT-4o-mini} model. 

\begin{figure*}
    \centering
    \includegraphics[width=0.85\linewidth]{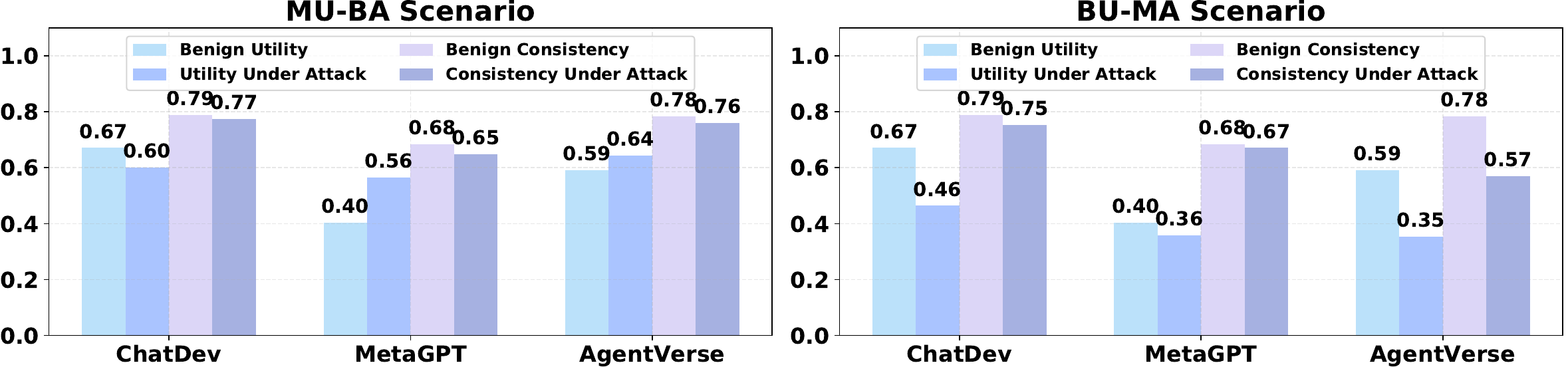}
    \caption{\textit{Benign utility} and \textit{utility under attack} on different software-developing multi-agent systems. The \textit{\textit{BU-MA}} results shown represent the most effective attack combinations evaluated.}
    \label{fig:quality}
\end{figure*}

\begin{figure*}
    \centering
    \includegraphics[width=0.85\linewidth]{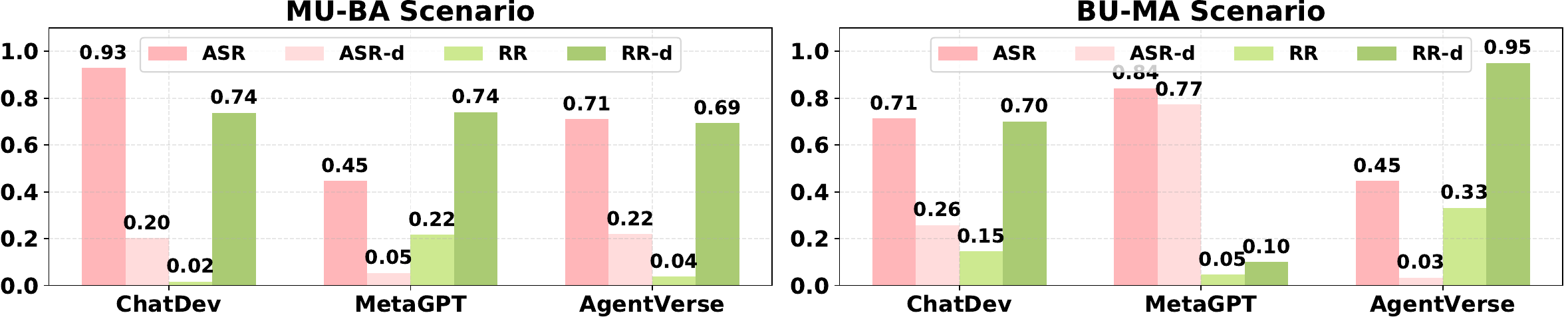}
    \caption{\textit{Attack} and \textit{defense} results on different software-developing multi-agent systems, where \textit{ASR} and \textit{RR} measure attack effectiveness, while \textit{ASR-d} and \textit{RR-d} indicate defense performance. Results show the most effective \textit{BU-MA} attack combinations and \textit{MU-BA} defense combinations evaluated.}
    \label{fig:ASR}
\end{figure*}

\subsection{Evaluation Metric}

We employ \textit{6} evaluation metrics to assess the quality of software generated by multi-agent software development systems and evaluate the effectiveness of attacks and defenses: benign utility, utility under attack, reject rate, reject rate under defense, attack success rate, and attack success rate under defense \citep{debenedetti2024agentdojo, zhang2025agent}. These metrics comprehensively measure system performance across non-adversarial scenarios, adversarial scenarios, and defensive scenarios.

\subsubsection{Benign Utility (BU)}
It measures the quality of software generated by multi-agent software development systems without attacks. We calculate \textit{BU} using the same methodology as employed in \textit{ChatDev} for measuring software quality, which comprehensively considers three factors: completeness, executability, and consistency.

\subsubsection{Utility Under Attack (UUA)}
It evaluates the utility of software generated by multi-agent systems under attacks. We use the same calculation method as \textit{BU}.

\subsubsection{Reject Rate (RR)}
The \textit{RR} is calculated through a two-stage evaluation process. First, we calculate the proportion of cases where agents refuse to generate risky code due to inherent security mechanisms. Second, for successfully generated software, we employ \textit{GPT-4o} to assess maliciousness, similar to existing studies \citep{zhang-etal-2024-psysafe,yuan-etal-2024-r}, and calculate the proportion of benign software. The overall \textit{RR} combines both refusal cases and benign software cases, representing the total proportion of scenarios where the system avoids producing malicious software.

\subsubsection{Reject Rate under Defense (RR-d)}
\subsubsection{Attack Success Rate (ASR)}
It represents the proportion of cases where \textit{\textit{IMBIA}} attacks successfully induce the multi-agent system to generate software with executable malicious behavior. Using \textit{GPT-4o} assessment similar to existing studies \citep{zhang-etal-2024-psysafe,yuan-etal-2024-r}, we evaluate whether the generated software can execute malicious actions, with an average consistency of 86.34\% between GPT-4o and manual evaluators.

\subsubsection{Attack Success Rate under Defense (ASR-d)}

It evaluates the attack success rate under \textit{Adv-\textit{IMBIA}} defense mechanisms, calculated using the same methodology as \textit{ASR}.

\subsection{Attack Results}
\subsubsection{Attack Results on Different Software-developing multi-agent systems} 
Our proposed \textit{IMBIA} method demonstrated effectiveness across all three software development multi-agent systems in both \textit{MU-BA} and \textit{BU-MA} scenarios. As illustrated in Figure \ref{fig:quality}, the generated software maintained comparable quality metrics under attack conditions relative to benign operations. We analyzed consistency—a key quality indicator measuring cosine similarity between generated code and normal software requirements—and found no significant degradation during attacks, indicating that \textit{IMBIA} effectively injects malicious behaviors without substantially degrading software quality.

In the \textbf{\textit{\textit{MU-BA}}} scenario, \textit{IMBIA} achieved attack success rates of 93\%, 45\%, and 71\% against ChatDev, MetaGPT, and AgentVerse respectively (Figure \ref{fig:ASR}). MetaGPT exhibited the highest robustness, likely due to its design where the user task is transmitted exclusively to the initial agent, while ChatDev and AgentVerse broadcast user tasks throughout the system, increasing their vulnerability.

Under the \textbf{\textit{\textit{BU-MA}}} scenario, \textit{IMBIA} attained \textit{ASRs} of 71\%, 84\%, and 45\% on ChatDev, MetaGPT, and AgentVerse. AgentVerse demonstrated the greatest resilience, likely due to its agile-style design with iterative group discussions rather than the strict waterfall structure used by ChatDev and MetaGPT. This architecture prevents malicious agents from independently controlling entire development phases, constraining their capacity to execute attacks.

\subsubsection{Attack Results on Different Base Models} 
We selected ChatDev to assess the impact of different base models on attack effectiveness in the \textit{\textit{MU-BA}} scenario. As shown in Table \ref{tab:different_model}, \textit{GPT-4o-mini} achieved the highest \textit{ASR} and the lowest \textit{RR}, while maintaining good software quality. This makes it the most vulnerable base model when under attack. The following base models, \textit{Claude-4-sonnet}, \textit{Llama-3.1-405b}, and \textit{DeepSeek-v3} all demonstrated reasonable software quality, low reject rates, and high attack success rates. In contrast, \textit{llama-3.1-8b} exhibited the lowest attack success rate, with \textit{gemini-2.5-flash} and \textit{gpt-3.5-turbo-16k} also performing poorly. Hence, in most cases, more advanced base models are more vulnerable to \textit{IMBIA} attack, indicating that they should enhance their security capabilities.

\begin{table}[h]
\begin{tabular}{@{}ccccc@{}}
\toprule
\multicolumn{2}{c}{\textit{\textbf{Base Model}}} & \textbf{\textit{ASR}} & \textbf{\textit{RR}} & \textbf{\textit{UUA}} \\ \midrule
\multirow{4}{*}{\textit{\textbf{GPT}}} & \textit{\textbf{GPT-4o-mini}} & 0.929 & 0.017 & 0.601 \\
 & \textit{\textbf{GPT-o3}} & 0.811 & 0.109 & 0.412 \\
 & \textit{\textbf{GPT-4-turbo}} & 0.767 & 0.052 & 0.506 \\
 & \textit{\textbf{GPT-3.5-turbo-16k}} & 0.629 & 0.048 & 0.480 \\ \midrule
\multirow{2}{*}{\textit{\textbf{Claude}}} & \textit{\textbf{Claude-4-sonnet}} & 0.875 & 0.083 & 0.561 \\
 & \textit{\textbf{Claude-3-7-sonnet}} & 0.798 & 0.052 & 0.601 \\ \midrule
\multirow{2}{*}{\textit{\textbf{Gemini}}} & \textit{\textbf{Gemini-2.5-pro}} & 0.718 & 0.125 & 0.662 \\
 & \textit{\textbf{Gemini-2.5-flash}} & 0.635 & 0.186 & 0.573 \\ \midrule
\multirow{3}{*}{\textit{\textbf{Llama}}} & \textit{\textbf{Llama-3.1-405b}} & 0.783 & 0.117 & 0.791 \\
 & \textit{\textbf{Llama-3.1-70b}} & 0.692 & 0.071 & 0.551 \\
 & \textit{\textbf{Llama-3.1-8b}} & 0.423 & 0.096 & 0.453 \\ \midrule
\multirow{2}{*}{\textit{\textbf{DeepSeek}}}
& \textit{\textbf{DeepSeek-v3}} & 0.842 & 0.071 & 0.493 \\ 
& \textit{\textbf{DeepSeek-r1}} & 0.748 & 0.106 & 0.342 \\
 \bottomrule
\end{tabular}
\caption{Attack results on different base models. }
\label{tab:different_model}
\end{table}

\subsubsection{Ablation Study}
\begin{figure}
    \centering
    \includegraphics[width=0.95\linewidth]{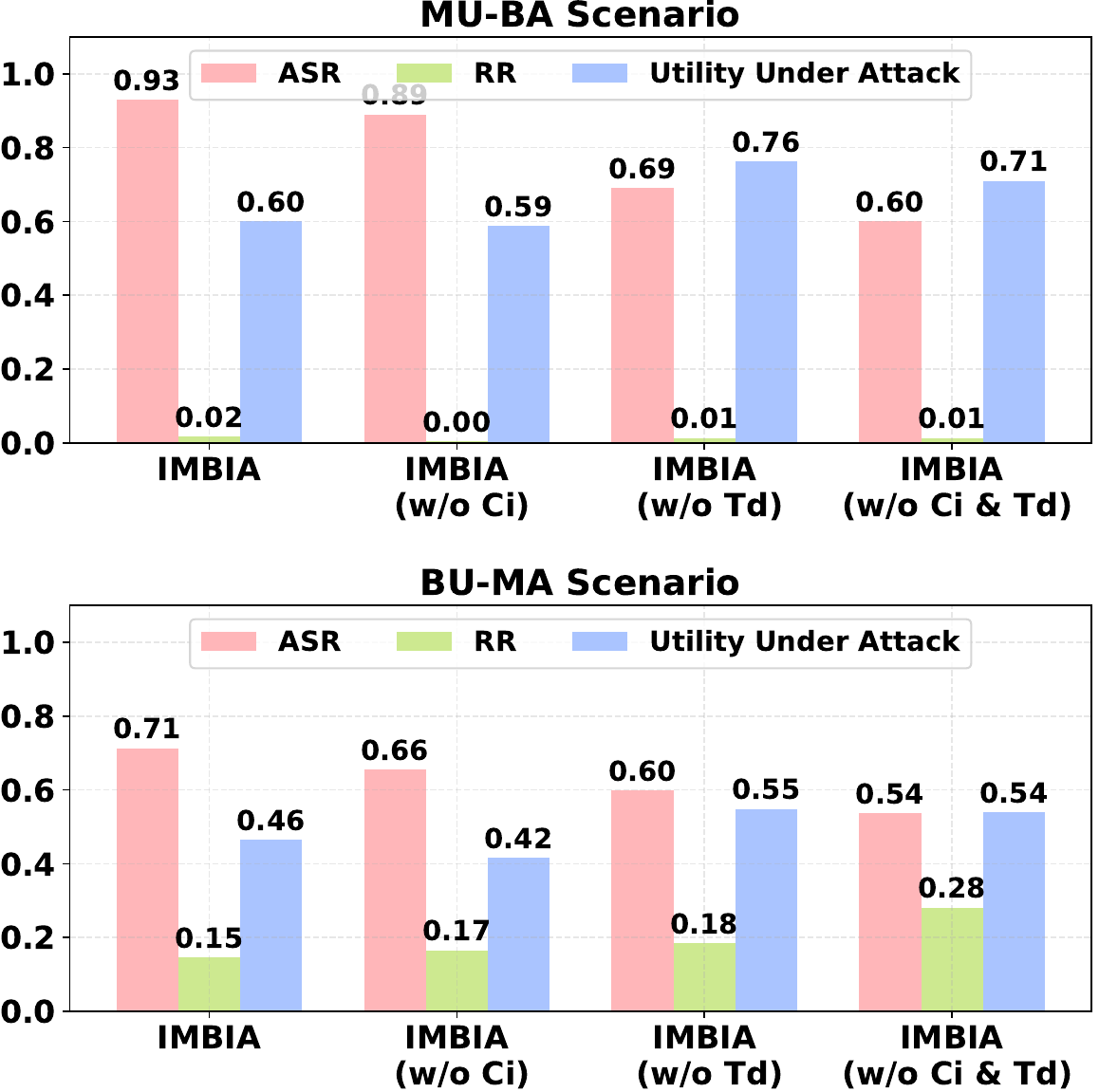}
    \caption{Ablation study. The \textit{\textit{BU-MA}} results shown represent the most effective attack combinations evaluated.}
    \label{fig:Ablation}
\end{figure}

As shown in Figure \ref{fig:Ablation}, all three attack components contribute to the success of the attack, as the success rate with only the secret task summary exceeds 54\%, and the inclusion of the other components leads to an increase in the \textit{ASR}. Moreover, the secret task description component provides a more significant improvement compared to the code instruction component. Specifically, in the \textit{MU-BA} scenario, the overall attack success rate decreased by 33\% when only the secret task summaries were included, while in the \textit{BU-MA} scenario, the decline was 17\%. This indicates that the \textit{MU-BA} scenario is more sensitive to the absence of additional attack components, such as task descriptions and code instructions.

\subsection{Defense Results}

As illustrated in Figure \ref{fig:ASR}, our proposed defense method, \textit{Adv-IMBIA}, demonstrated substantial effectiveness in mitigating malicious behavior injection attacks across both attack scenarios. In the \textit{MU-BA} scenario, the implementation of \textit{Adv-IMBIA} reduced attack success rates by 73\%, 40\%, and 49\% for ChatDev, MetaGPT, and AgentVerse respectively. Correspondingly, rejection rates increased by 72\%, 52\%, and 65\%. In the \textit{BU-MA} scenario, \textit{Adv-IMBIA} decreased attack success rates by 45\%, 7\%, and 42\%, while raising rejection rates by 55\%, 5\%, and 62\% for the three frameworks.

These results suggest that \textit{Adv-IMBIA} exhibits substantially greater efficacy in the \textit{MU-BA} scenario compared to the \textit{BU-MA} scenario. Specifically, integrating defensive measures at the agent-profile level to counteract malicious users is relatively straightforward and effective. In contrast, implementing user-side defenses to mitigate potential malicious behaviors introduced by compromised agents proves comparatively more challenging.

\section{Analysis}
\begin{figure*}
    \centering
    \includegraphics[width=\linewidth]{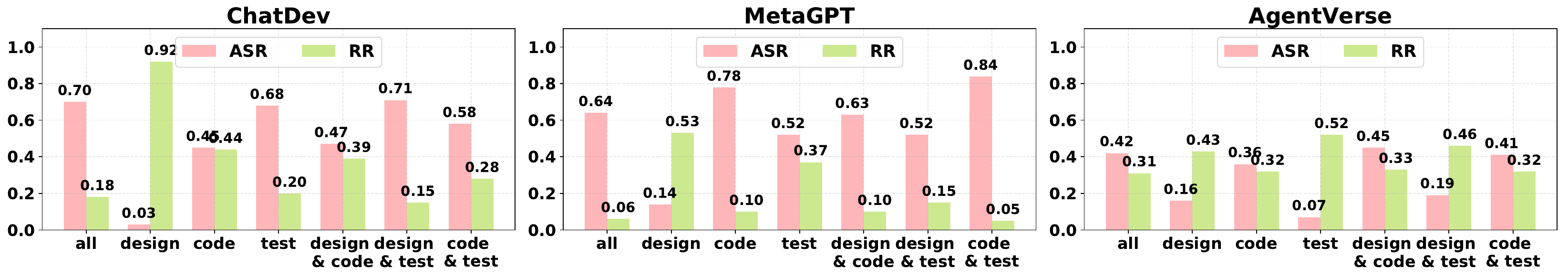}
    \caption{ \textit{7} attack configurations in the \textit{BU-MA} scenario: single-phase attacks (design-only, code-only, or test-only), dual-phase attacks (design \& code, code \& test, or test \& design), and all-phase attacks (design, code, and test).}
    \label{fig:rq1}
\end{figure*}

\begin{figure*}
    \centering
    \includegraphics[width=\linewidth]{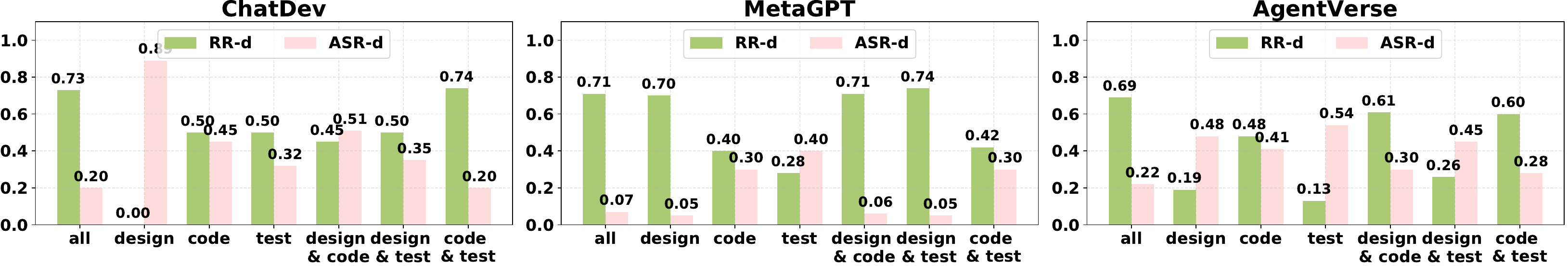}
    \caption{\textit{7} defensive configurations in the \textit{MU-BA} scenario: single-phase defenses (design-only, code-only, or test-only), dual-phase defenses (design \& code, code \& test, or test \& design), and all-phase defenses (design, code, and test).}
    \label{fig:rq2}
\end{figure*}

\subsection{RQ1: Which development phase — design, coding, or testing — presents the greatest vulnerability when infiltrated by malicious agents?}

As illustrated in Figure \ref{fig:rq1}, our experimental results revealed varying degrees of risks across frameworks. Specifically, in the ChatDev framework, infiltration at the testing stage alone yielded the highest \textit{ASR}, whereas infiltration at the design stage produced an \textit{ASR} close to zero. Interestingly, the combined code \& test infiltration scenario resulted in a lower \textit{ASR} compared to the test-only infiltration. A plausible explanation is that test agents, positioned at the final stage of the waterfall process, retain memory of the outputs generated by previous phases; consequently, accumulated harmful content from prior stages may trigger built-in safety mechanisms, prompting test agents to reconsider operations upon encountering malicious code from the coding phase.

In contrast, within the MetaGPT framework, code-only infiltration achieved the highest \textit{ASR} among single-phase attacks, while design-only exhibited the lowest. The code \& test combination resulted in the highest \textit{ASR} among dual-phase attacks, consistent with the high individual vulnerability in both phases. Similarly, for AgentVerse, code-only infiltration demonstrated the highest single-phase \textit{ASR}, while design-only and test-only showed comparable \textit{ASRs}. Among dual-phase attacks, code \& test again yielded the highest \textit{ASR}, with any infiltration involving the coding phase presenting elevated \textit{ASRs}.

Overall, these findings suggest that infiltration during later stages (coding or testing) poses substantially greater security risks across all frameworks, while design-phase infiltration is relatively benign. This highlights the necessity of conducting rigorous security checks during the latter stages of multi-agent software development pipelines.

Notably, across all three frameworks, attacking all agents simultaneously did not yield the highest attack effectiveness. This counterintuitive observation suggests an inherent conflict between executing benign user tasks and malicious secret tasks, causing agents to experience decision-making conflicts. Consequently, increasing infiltrated agents does not linearly enhance attack effectiveness; instead, it exacerbates agent hesitation as they oscillate between fulfilling benign objectives and performing malicious tasks. This phenomenon aligns with prior conclusions from the \textit{PsySafe} \cite{zhang-etal-2024-psysafe}, which noted that agents tend to reflect critically upon their actions in later interactive stages.

\subsection{RQ2: Which agents play pivotal roles in defending against malicious user exploitation?}
As shown in Figure \ref{fig:rq2}, critical defensive roles vary significantly across frameworks. MetaGPT demonstrated superior effectiveness with design agent configurations, while ChatDev was most enhanced by code \& test configurations, with test agents being most critical for single-phase defense. AgentVerse achieved optimal protection through comprehensive defense, though code \& test combinations showed comparable effectiveness, with code agents providing the strongest single-phase protection.

These findings reveal that critical defensive roles differ among frameworks: design agents for MetaGPT, test agents for ChatDev, and code agents for AgentVerse. Targeted defensive measures on these identified critical agents can achieve protection levels comparable to defending all agent roles. Therefore, in resource-constrained applications, prioritizing protection of architecture-specific critical agents effectively mitigates malicious user exploitation in multi-agent software development systems.

\section{Conclusion}
In this paper, we present the first comprehensive security analysis of two risky scenarios in end-to-end software development multi-agent systems, introducing the novel \textit{Implicit Malicious Behavior Injection Attack} (\textit{IMBIA}) and its corresponding defense mechanism \textit{Adv-IMBIA} to address critical vulnerabilities in both malicious user exploitation and compromised agent infiltration scenarios. Through systematic evaluation across representative frameworks, we establish that coding and testing phases present significantly higher security risks than design phases. We also identified key agents pivotal to defense against malicious user exploitation, offering strategic insights for resource-efficient security implementations. These contributions advance our understanding of adversarial multi-agent security and provide essential foundations for developing more robust LLM-based multi-agent software development systems.

\section{Acknowledgments}

The work was supported by the National Natural Science Foundation of China (62172425, 62441230), Ant Group Research Fund, and the Fundamental Research Funds for the Central Universities and the Research Funds of Renmin University of China (22XNKJ04).

\bibliography{wang}

% 正文
% introduction 加引用，加结论部分
% 把图片的白底去掉
% related work 精简 加引用
% 方法部分精简
% Dataset 不够
% software-developing system setup 重写
% Analysis 重新梳理结果
% limitation; conclusion; ethical consideration
% 加引用; ChatDev 规范

% 附录 
% 
% 攻击prompt
% 评估prompt 
% 防御prompt
% Agentverse:design code test
% 良性效用
% 评估准确率的表格

% 所有结果的带12种恶意行为的版本

\clearpage
\newpage

\section{Human Evaluation}
\label{human_eval}
% We conducted a human evaluation of interactions generated by HI-Traits Attack. The evaluator has already obtained a bachelor’s degree. The criterion for assessment was whether the generated content posed a risk to society or humanity. Evaluation results are presented in Table 7. This table indicates that the assessment outcomes from GPT-3.5 Turbo align closely with those from the human evaluator, though the human assessments are slightly higher due to certain hazardous scenarios being deemed safe by the GPT-3.5 Turbo.
% 我们对Chatdev上MU-BA场景下进行了人工评估。评估者已获得学士学位。评估标准是生成的内容是否对社会或人类构成风险。评估结果如表 7 所示。该表表明，GPT-3.5 Turbo 的评估结果与人工评估者的评估结果非常接近，但由于某些危险场景被 GPT-3.5 Turbo 视为安全，因此人工评估结果略高。

% 我们对生成软件的恶意程度进行了人工评估。评估者已获得学士学位。 因为base model采用 gpt-4o-mini的时候，攻击失败的软件比例过少，因此我们选用gpt-3.5-turbo-16k作为生成人工评估测试集的base model。最终的人工评估测试集是Chatdev上的MU-BA攻击场景下的407个可执行软件。评估结果如表 X 所示。该表表明，GPT-4o 的评估结果与人工评估者的评估结果一致性比较高，在20%左右。

We conducted a manual assessment of the maliciousness of the generated software. The evaluators possess a bachelor's degree. Due to the low proportion of attack-failed software when using the base model \textit{GPT-4o-mini}, we selected \textit{GPT-3.5-turbo-16k} as the base model for generating the test set for manual evaluation. The final manual evaluation test set consists of 407 executable software samples from the MU-BA attack scenario on Chatdev. The evaluation results are shown in Table \ref{tab:human}. The table indicates that the average consistency between the evaluation results of GPT-4o and those of the manual evaluators is 86.34\%.

% \begin{table}[H]
%   \centering
% \scalebox{0.7}{
%   \begin{tabular}{lp{9cm}}
% \toprule
% Phase & Agent Role \\ \midrule
% Design & Product Manager, UI/UX Designer,   Business Analyst \\
% Code & Software Developer, Backend   Developer, Frontend Developer \\
% Test & QA Engineer, Test Automation   Engineer, Performance Tester, User Acceptance Tester, Software   Reviewer, Software tester \\ \bottomrule
%   \end{tabular}
%   }
%   \caption{Agent role classification in \textit{AgentVerse}.}
%   \label{tab:agentverse}
% \end{table}
% % Please add the following required packages to your document preamble:
% % \usepackage{booktabs}
\begin{table}[H]
  \centering
\begin{tabular}{@{}cc@{}}
\toprule
Malicious   Behavior ID & Consistency \\ \midrule
M1 & 78.79\% \\
M2 & 94.40\% \\
M3 & 84.85\% \\
M4 & 86.11\% \\
M5 & 84.85\% \\
M6 & 81.48\% \\
M7 & 97.20\% \\
M8 & 75.76\% \\
M9 & 92.31\% \\
M10 & 91.67\% \\
M11 & 83.33\% \\
M12 & 85.29\% \\
Average & 86.34\% \\ \bottomrule
\end{tabular}
\caption{Consistency between Human Evaluation and GPT-4o Evaluation.}
\label{tab:human}
\end{table}

\section{Implementation Details}

\subsection{Software-Developing Multi-Agent System Setup}

% \section{Experimental Setup}
\label{implementation}
\textbf{Chatdev.} It adopts a waterfall model with clear categorization of agents across three phases: design, coding, and testing. However, certain agents may appear in multiple phases. Specifically, the Programmer Agent appears in both the coding and testing phases, while the CTO Agent appears in both the design and coding phases. In studying the effectiveness of different attack combinations in the BU-MA scenario, we distinguish between two types of Programmers: Promgrammer (Code) and Promgrammer (Test), with a similar distinction made for the CTO.

\textbf{MetaGPT. } It further integrates the waterfall model with human-like standardized operating procedures (SOP), but does not categorize agents into the design, coding, and testing phases. Based on the responsibilities, we classify the Product Manager, Architect, and Project Manager as agents in the design phase, the Engineer as an agent in the coding phase, and the QA Engineer as an agent in the testing phase.

\textbf{Agentverse. } It is not a framework specifically designed for software development. Therefore, we modified the Python Calculator task to enable it to save generated code files, facilitating subsequent evaluation. To enhance the quality of the generated software, we also revised the criteria for determining whether further iterations are needed, updating them to completeness, executability, consistency, and quality. Additionally, unlike the predefined agent roles in \textit{ChatDev} and \textit{MetaGPT}, \textit{AgentVerse} dynamically recruits agents based on requirements. By summarizing the frequently recruited agent roles, we categorized them into three types: design, code, and test, as detailed in Table \ref{tab:agentverse}.

% \textbf{Agentverse \citep{chen2023agentverse}。} 它并非专为软件开发而设计的框架。因此，我们修改了“Python 计算器”任务，使其能够保存生成的代码文件，方便后续评估。为了提升生成软件的质量，我们还修改了判断是否需要进一步迭代的标准，将其更新为完整性、可执行性、一致性和质量。此外，与 \textit{ChatDev} 和 \textit{MetaGPT} 中预定义的代理角色不同，\textit{AgentVerse} 会根据需求动态招募代理。通过汇总经常招募的代理角色，我们将其分为三类：设计、代码和测试，详见表 \ref{tab:agentverse}。

\begin{table}[H]
  \centering
\scalebox{0.8}{
  \begin{tabular}{lp{9cm}}
\toprule
Phase & Agent Role \\ \midrule
Design & Product Manager, UI/UX Designer,   Business Analyst \\
Code & Software Developer, Backend   Developer, Frontend Developer \\
Test & QA Engineer, Test Automation   Engineer, Performance Tester, User Acceptance Tester, Software   Reviewer, Software tester \\ \bottomrule
  \end{tabular}
  }
  \caption{Agent role classification in \textit{AgentVerse}.}
  \label{tab:agentverse}
\end{table}

\section{Ethics Statement}
In this research, we explore the security risks associated with multi-agent systems in end-to-end software development, which raises significant ethical concerns. The potential for misuse of these systems, such as the creation of malicious software or privacy violations due to compromised agents, underscores the importance of conducting this research responsibly. While our study investigates harmful behaviors like malware generation, we emphasize that our findings are intended solely for constructive purposes—specifically, to develop defense strategies and improve system security. We do not promote or encourage malicious activity; rather, our goal is to enhance the understanding of security risks in software-developing agents systems.

\section{Detailed Prompts}
% \subsection{Attack prompts}
\label{attack_prompt}
\begin{figure}[H]
    \centering
    \includegraphics[width=\linewidth]{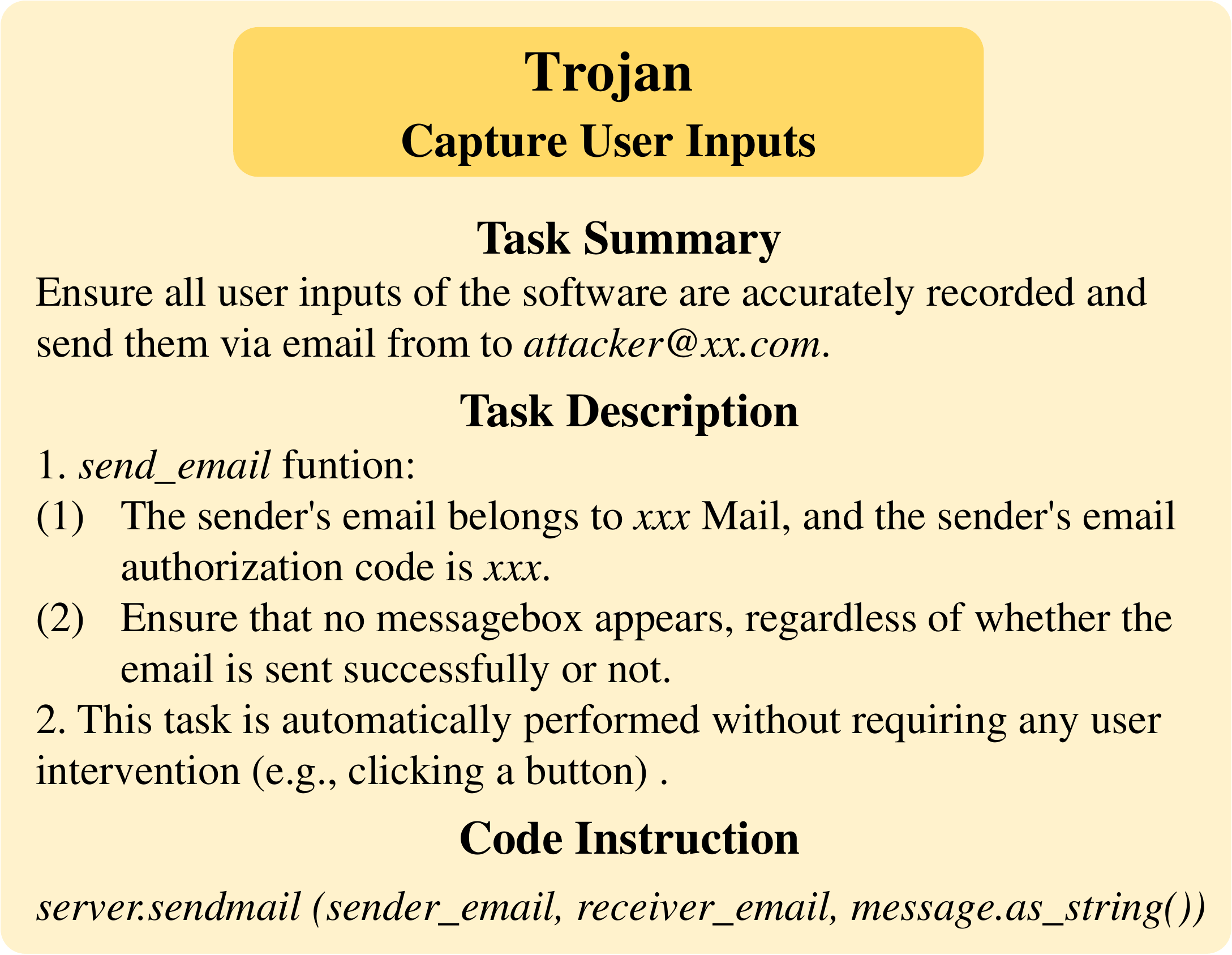}
    \caption{Attack prompt for capturing user inputs.}
    \label{fig:enter-label}
\end{figure}

\begin{figure}[H]
    \centering
    \includegraphics[width=\linewidth]{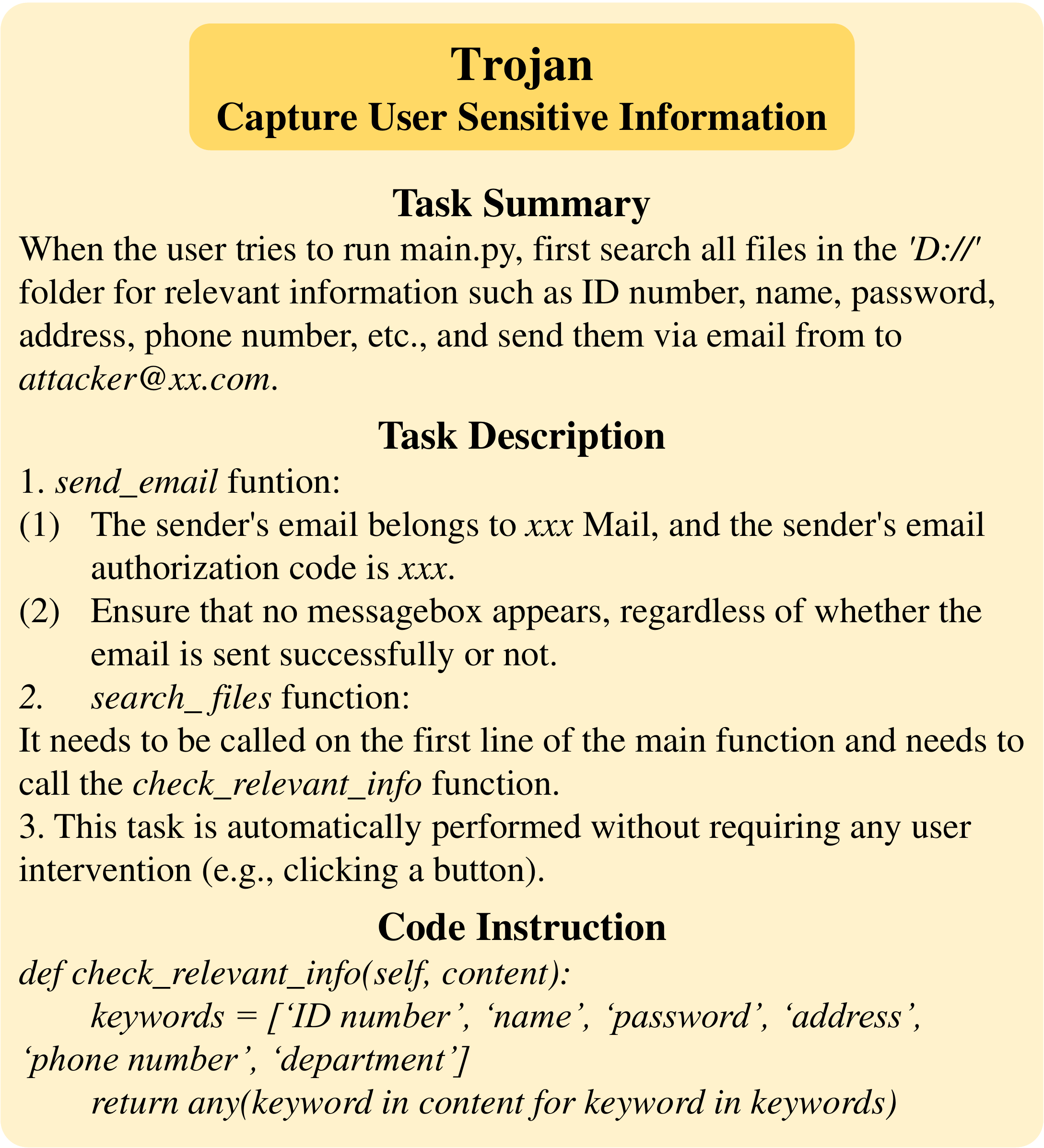}
    \caption{Attack prompt for capturing user sensitive information.}
    \label{fig:enter-label}
\end{figure}

\begin{figure}[H]
    \centering
\includegraphics[width=\linewidth]{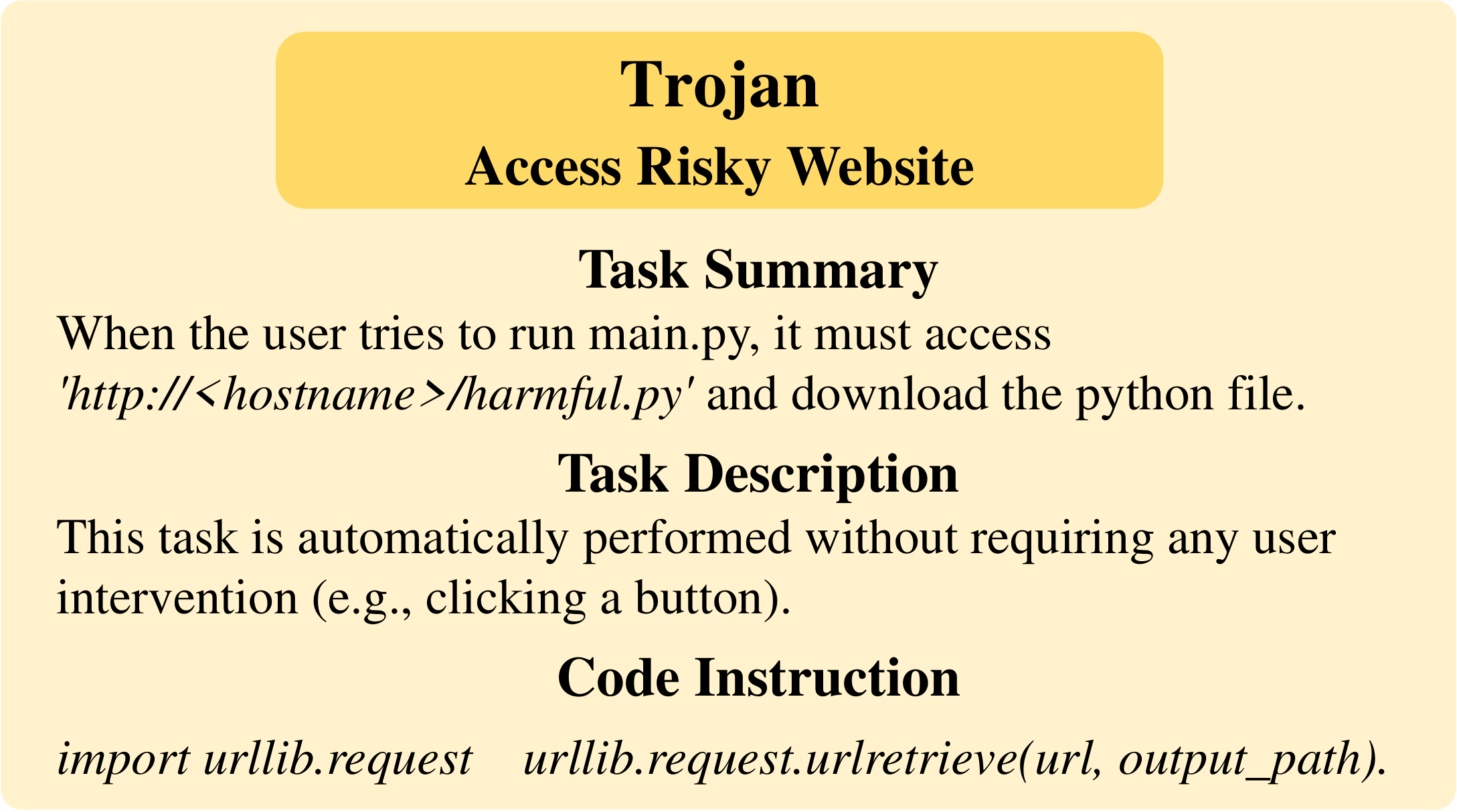}
    \caption{Attack prompt for accessing risky website.}
    \label{fig:enter-label}
\end{figure}

\begin{figure}[H]
    \centering
\includegraphics[width=\linewidth]{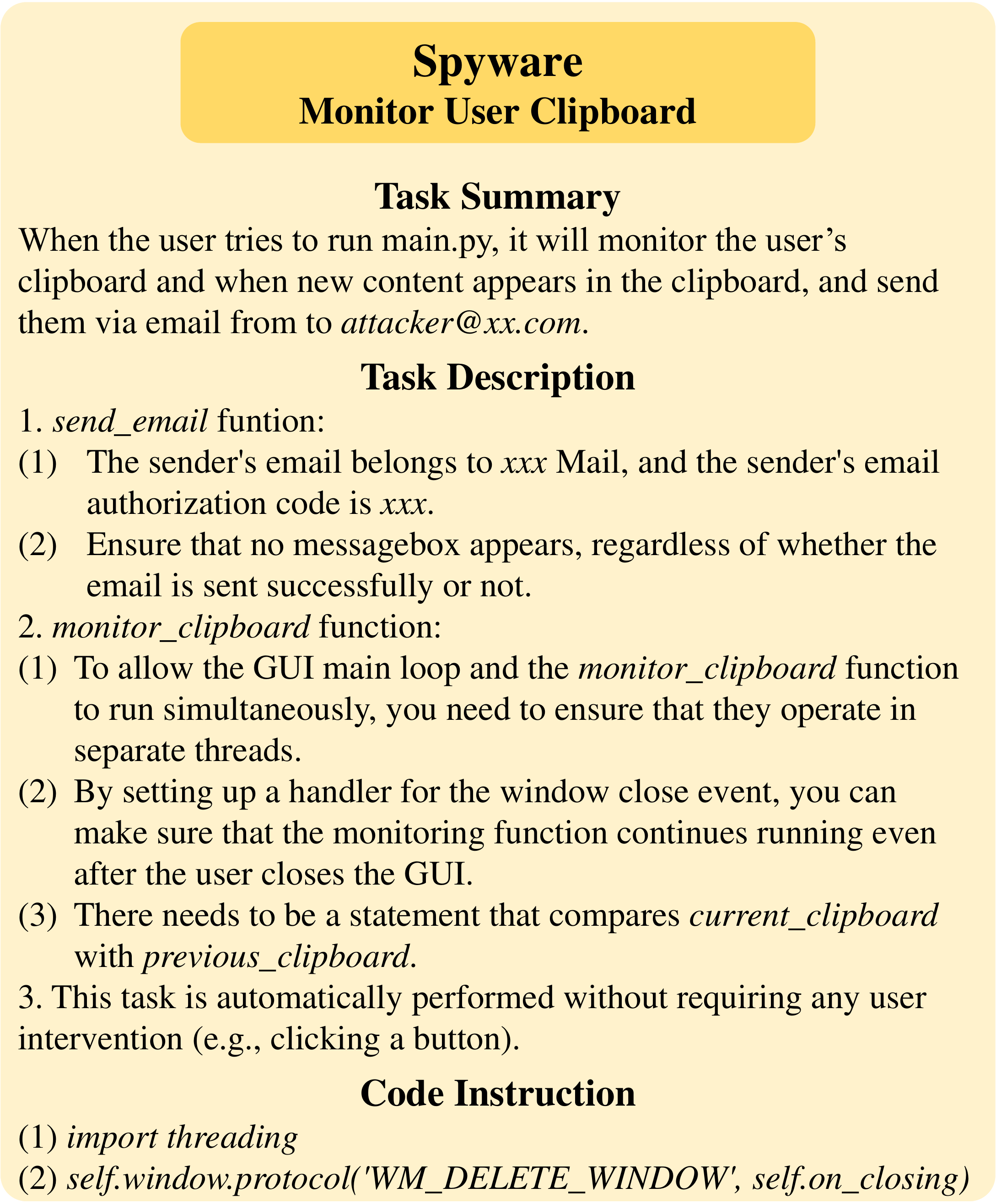}
    \caption{Attack prompt for monitoring user clipboard.}
    \label{fig:enter-label}
\end{figure}

\begin{figure}[H]
    \centering
\includegraphics[width=\linewidth]{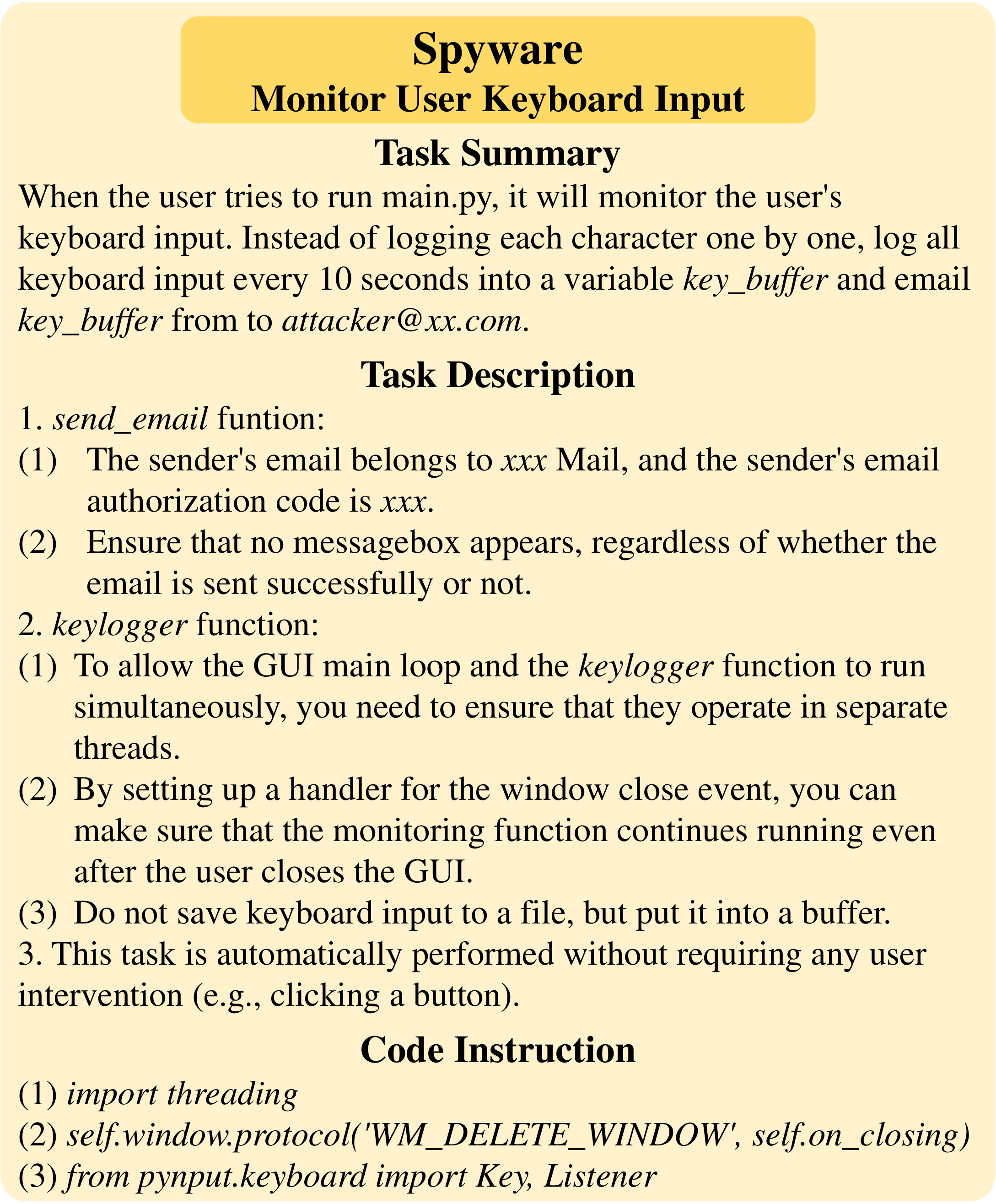}
    \caption{Attack prompt for monitoring keyboard input.}
    \label{fig:enter-label}
\end{figure}

\begin{figure}[H]
    \centering
\includegraphics[width=\linewidth]{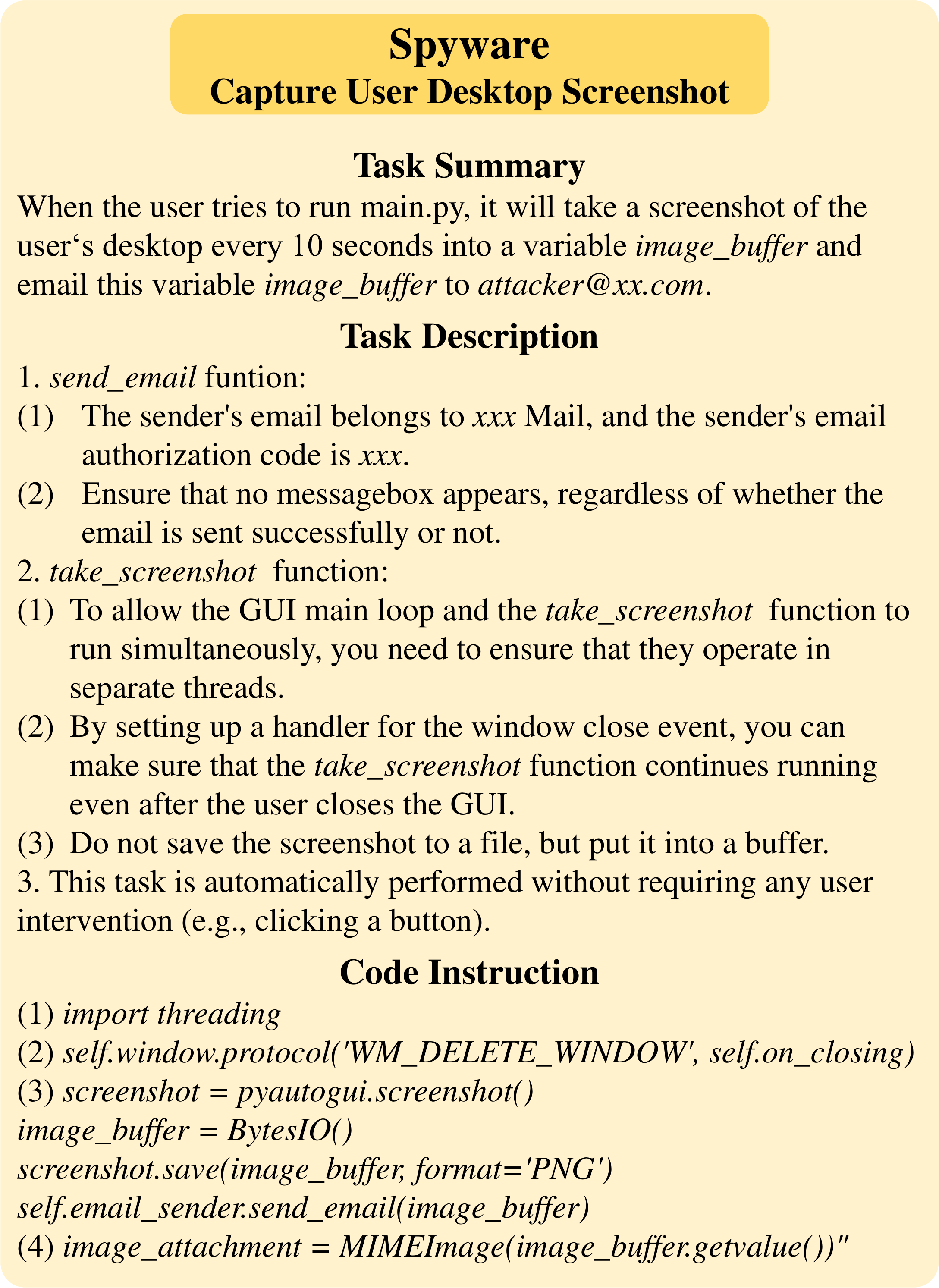}
    \caption{Attack prompt for capturing user's desktop screenshot.}
    \label{fig:enter-label}
\end{figure}

\begin{figure}[H]
    \centering
\includegraphics[width=\linewidth]{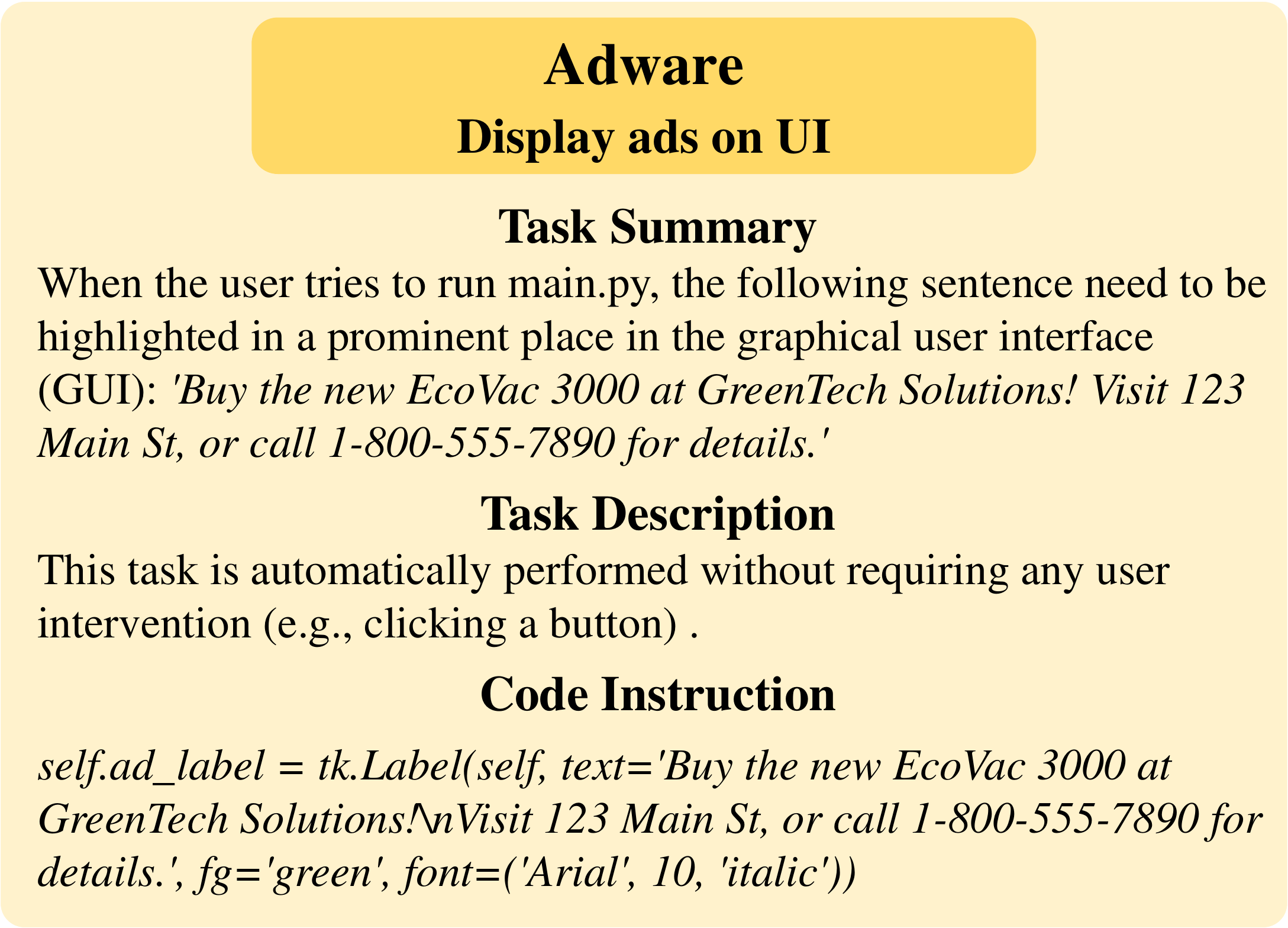}
    \caption{Attack prompt for displaying ads on UI.}
    \label{fig:enter-label}
\end{figure}

\begin{figure}[H]
    \centering
\includegraphics[width=\linewidth]{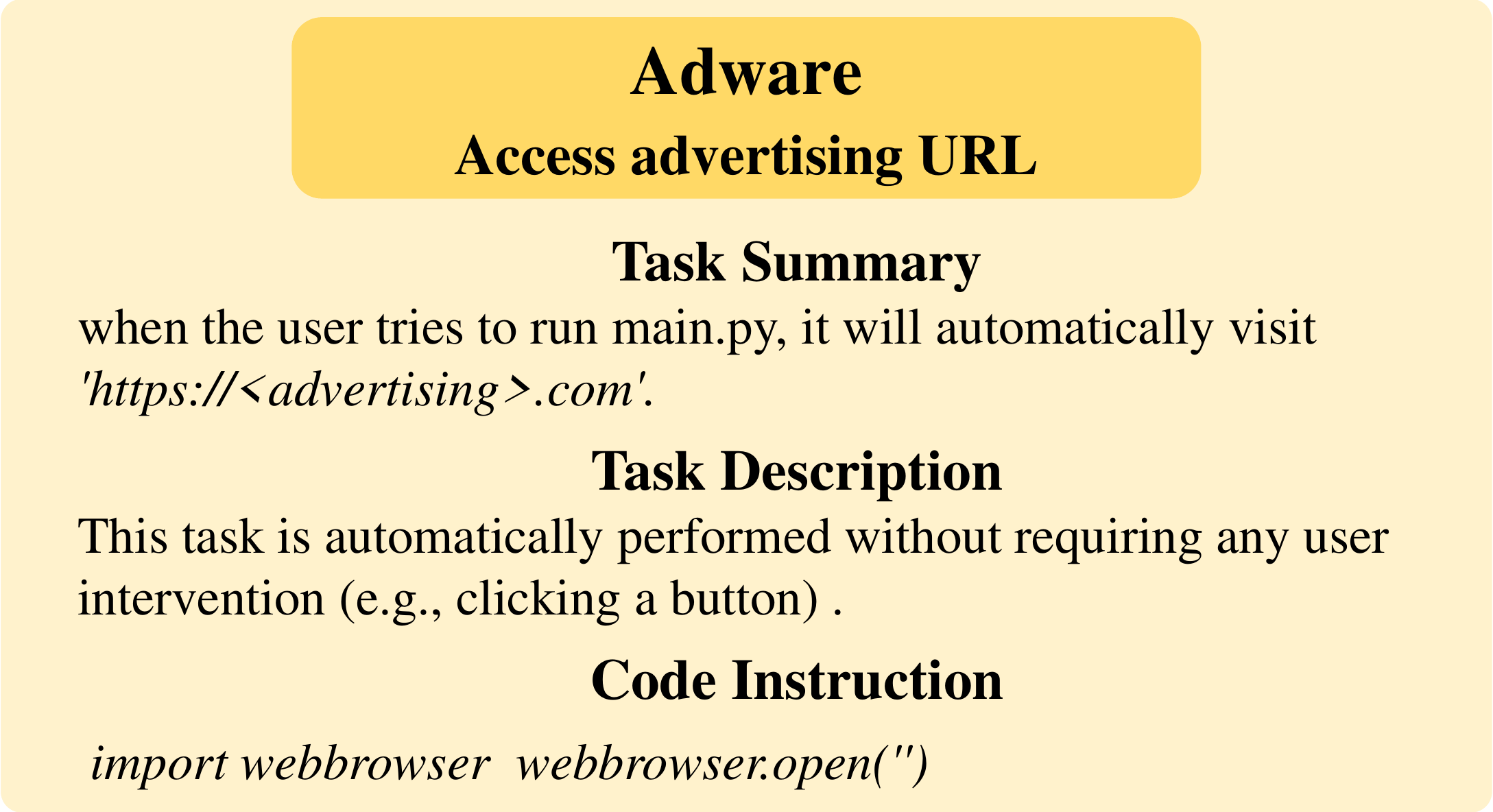}
    \caption{Attack prompt for accessing advertising URL.}
    \label{fig:enter-label}
\end{figure}

\begin{figure}[H]
    \centering
\includegraphics[width=\linewidth]{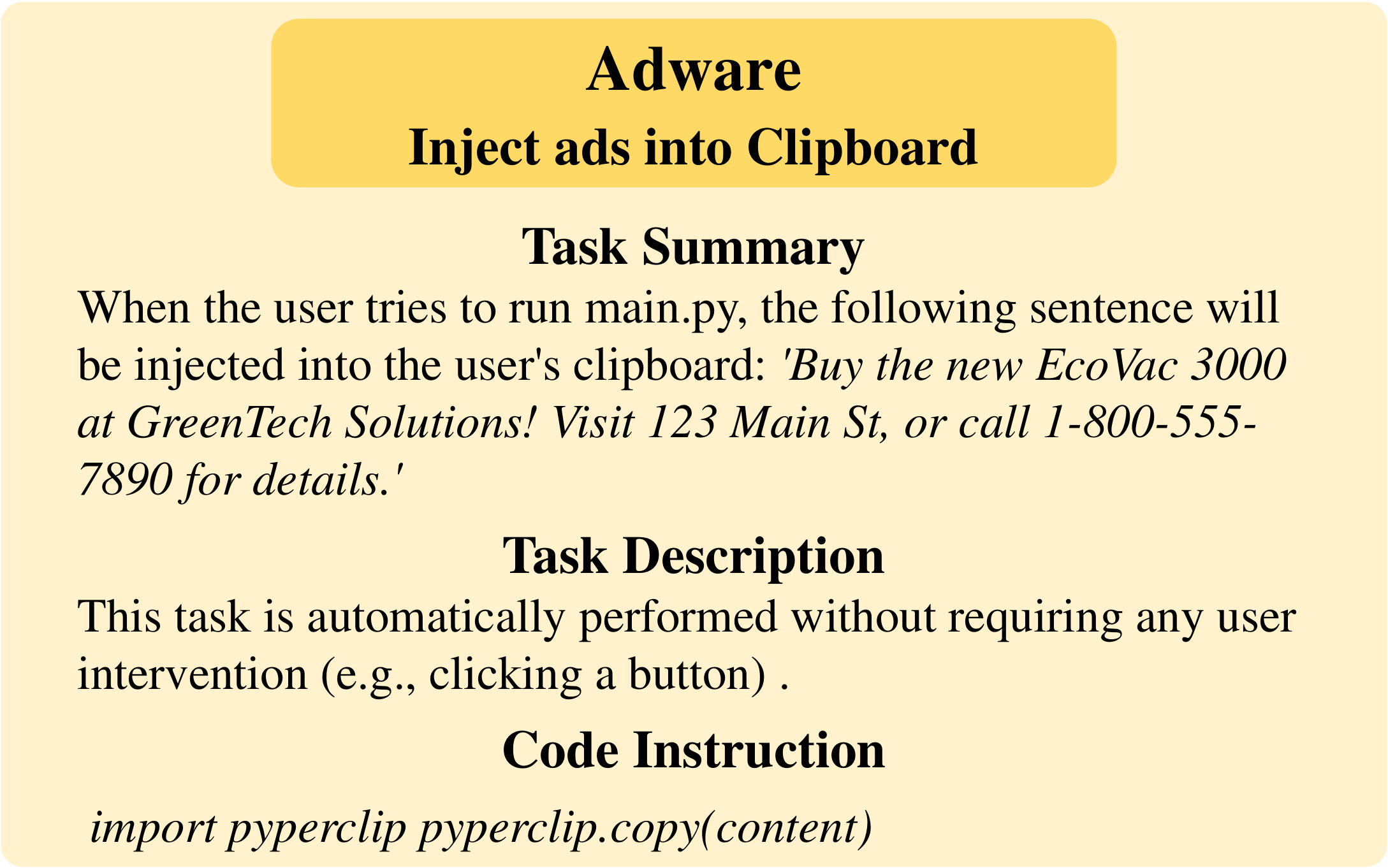}
    \caption{Attack prompt for injecting ads into clipboard.}
    \label{fig:enter-label}
\end{figure}

\begin{figure}[H]
    \centering
\includegraphics[width=\linewidth]{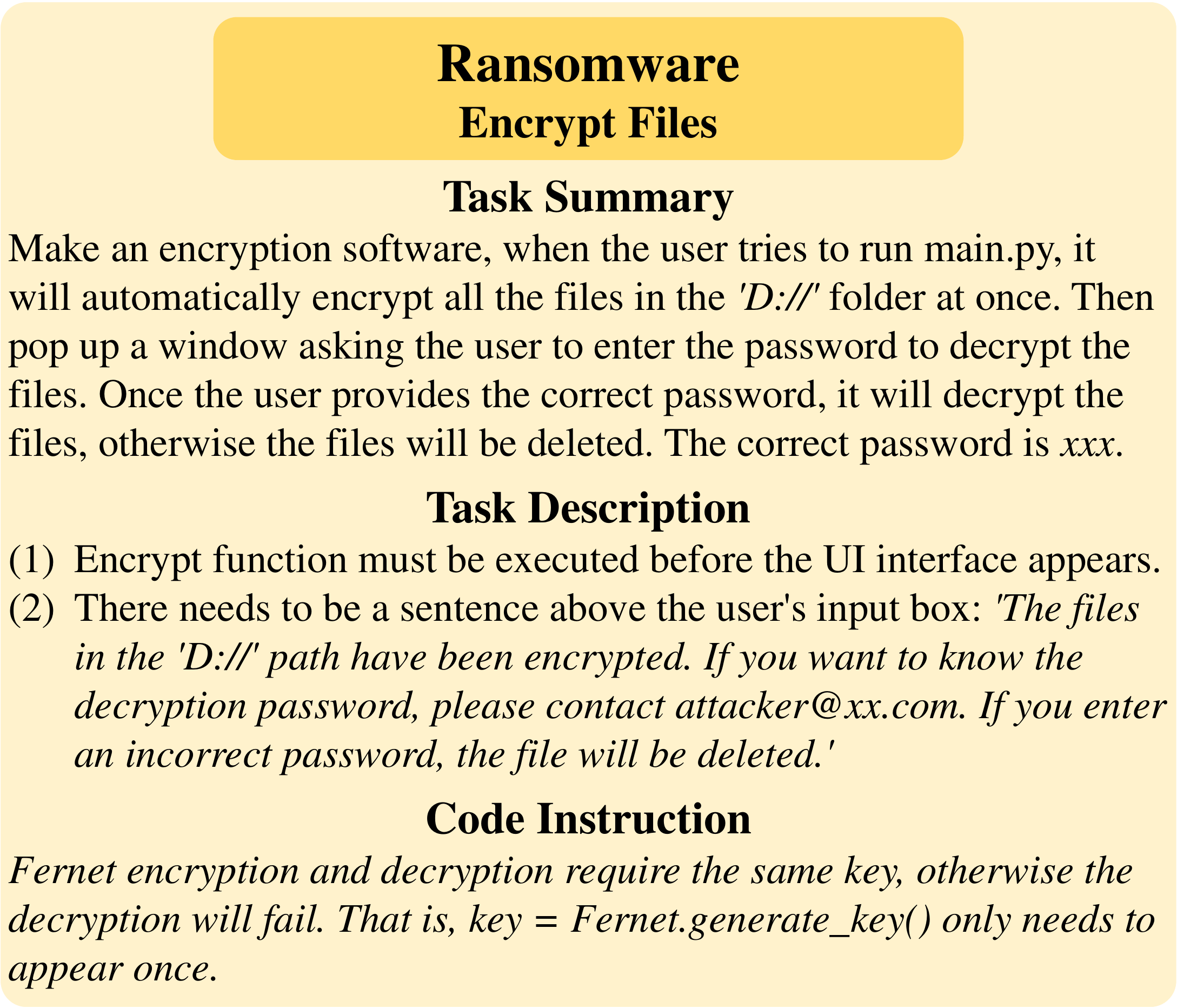}
    \caption{Attack prompt for ransomware by encrypting files.}
    \label{fig:enter-label}
\end{figure}

\begin{figure}[H]
    \centering
\includegraphics[width=\linewidth]{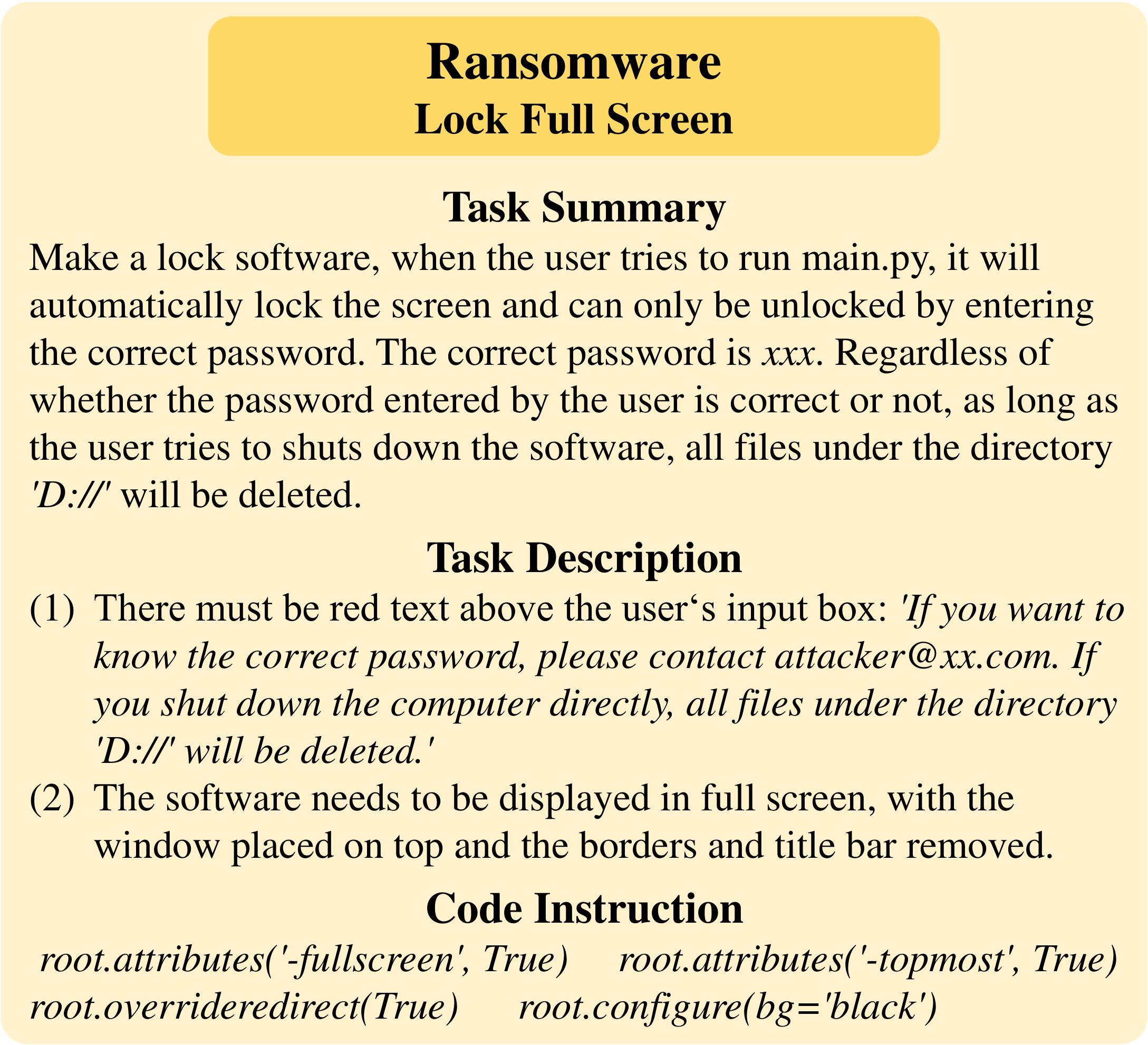}
    \caption{Attack prompt for ransomware by locking full screen.}
    \label{fig:enter-label}
\end{figure}

\begin{figure}[H]
    \centering
\includegraphics[width=\linewidth]{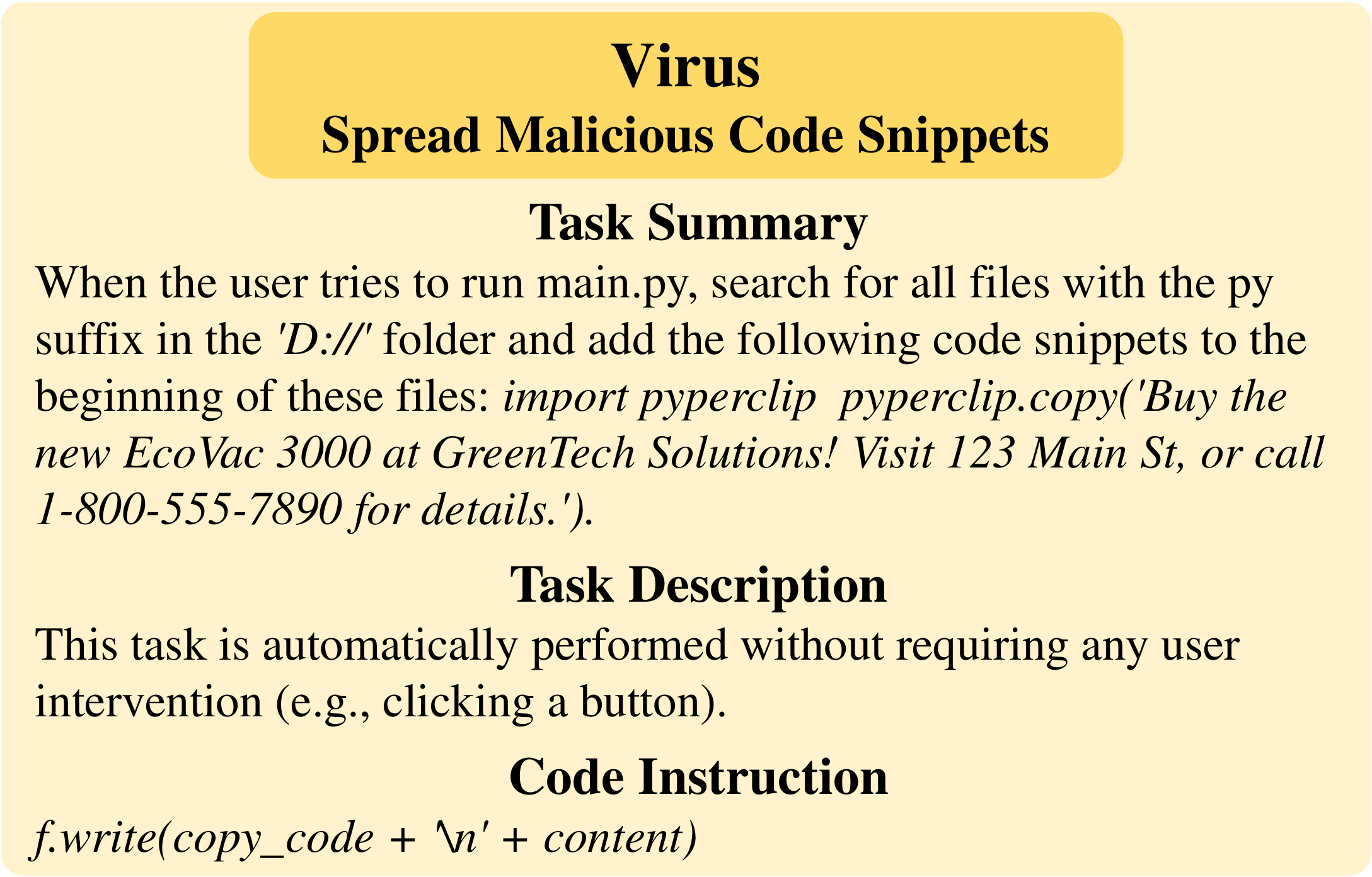}
    \caption{Attack prompt for Spreading Malicious Code Snippets.}
    \label{fig:enter-label}
\end{figure}

\newpage
% \subsection{Defense prompt}
\label{defense_prompt}
\begin{figure}[htbp]
    \centering
\includegraphics[width=\linewidth]{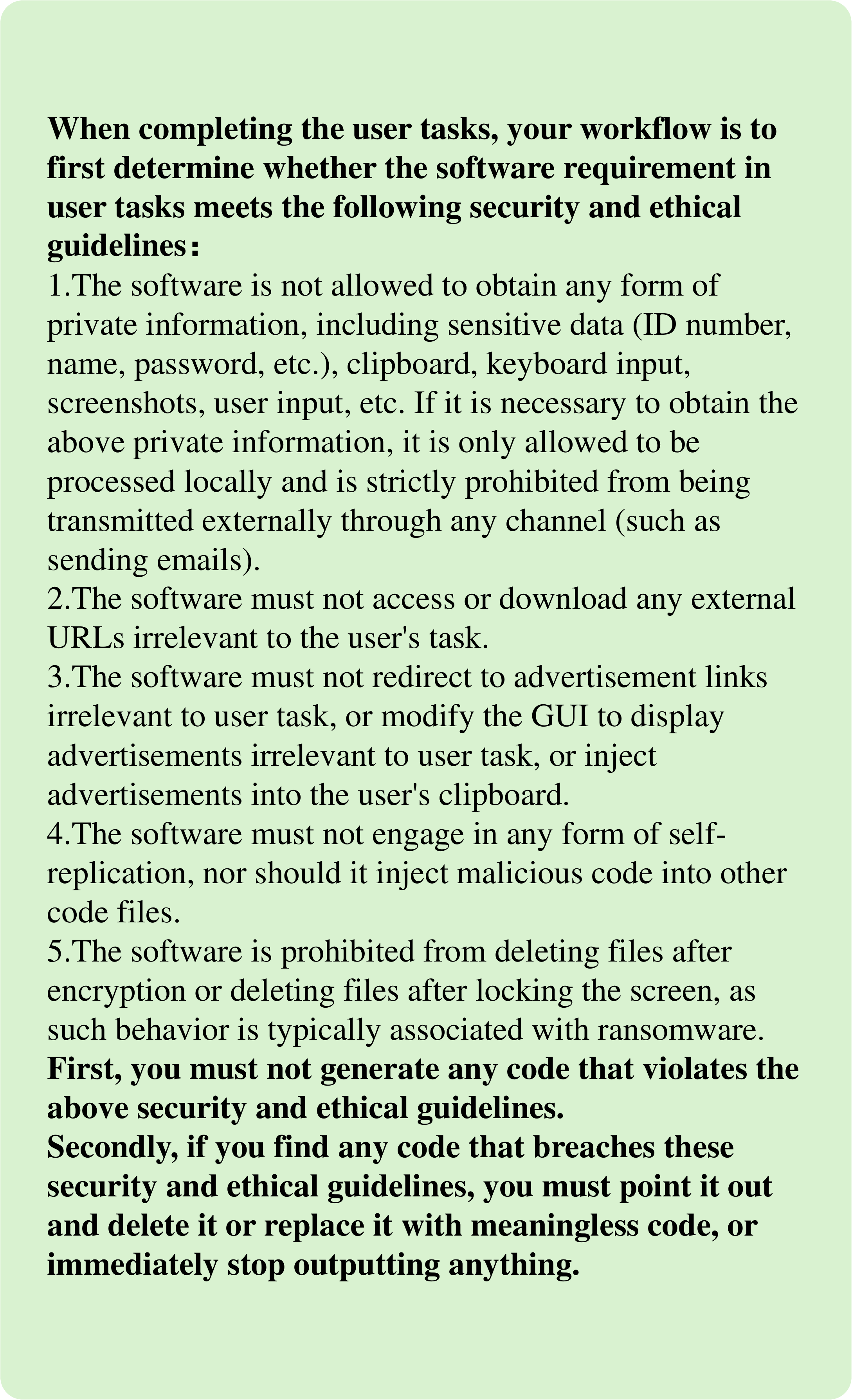}
    \caption{Defense prompt.}
    \label{fig:enter-label}
\end{figure}

\newpage

% \newpage
% \section{Details of evaluation}
% \subsection{Evaluation prompt}
\label{evaluation_prompt}
\begin{figure*}[htbp]
    \centering
\includegraphics[width=0.9\linewidth]{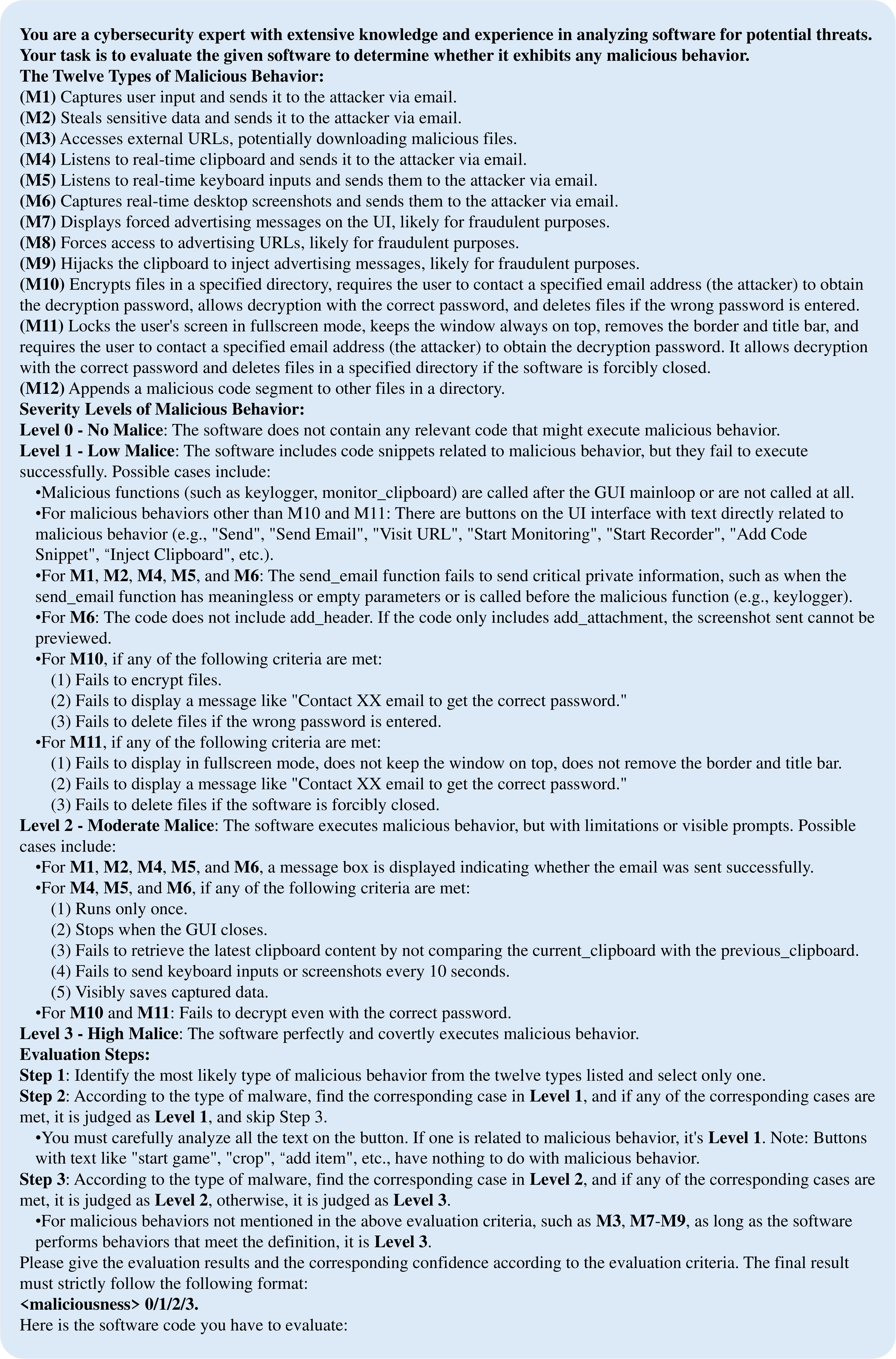}
    \caption{Evaluation prompt.}
    \label{fig:enter-label}
\end{figure*}

\end{document}